\pdfoutput=1

\documentclass[sigconf]{acmart}

\AtBeginDocument{%
  \providecommand\BibTeX{{%
    \normalfont B\kern-0.5em{\scshape i\kern-0.25em b}\kern-0.8em\TeX}}}

\copyrightyear{2022}
\acmYear{2022}
\setcopyright{rightsretained}
\acmConference[CHI '22]{CHI Conference on Human Factors in Computing Systems}{April 29-May 5, 2022}{New Orleans, LA, USA}
\acmBooktitle{CHI Conference on Human Factors in Computing Systems (CHI '22), April 29-May 5, 2022, New Orleans, LA, USA}
\acmDOI{10.1145/3491102.3501972}
\acmISBN{978-1-4503-9157-3/22/04}

\newcommand{\scstart}[1]{\vspace{.3mm} \noindent{\textsc{#1:}}}

\newcommand{\change}[1]{\textcolor{black}{#1}}

\usepackage{subfigure}

\acmSubmissionID{4814}

\begin{document}

\title{How do you Converse with an Analytical Chatbot? \\ Revisiting Gricean Maxims for Designing Analytical Conversational Behavior}


\author{Vidya Setlur}
\affiliation{%
  \institution{Tableau Research}
  \streetaddress{260 California Ave STE 300, Palo Alto}
  \city{Palo Alto}
  \state{CA}
  \postcode{94306}
  \country{USA}}
\email{vsetlur@tableau.com}

\author{Melanie Tory}
\affiliation{%
  \institution{The Roux Institute, Northeastern University}
   \streetaddress{100 Fore St.}
  \city{Portland}
  \state{ME}
  \postcode{04101}
  \country{USA}}
  \email{m.tory@northeastern.edu}


\begin{abstract}
Chatbots have garnered interest as conversational interfaces for a variety of tasks. While general design guidelines exist for chatbot interfaces, little work explores analytical chatbots that support conversing with data. We explore Gricean Maxims to help inform the basic design of effective conversational interaction. We also draw inspiration from natural language interfaces for data exploration to support ambiguity and intent handling. We ran Wizard of Oz studies with 30 participants to evaluate user expectations for text and voice chatbot design variants. Results identified preferences for intent interpretation and revealed variations in user expectations based on the interface affordances. We subsequently conducted an exploratory analysis of three analytical chatbot systems (text + chart, voice + chart, voice-only) that implement these preferred design variants. Empirical evidence from a second 30-participant study informs implications specific to data-driven conversation such as interpreting intent, data orientation, and establishing trust through appropriate system responses.
\end{abstract}



\begin{teaserfigure}
    \centering
  \includegraphics[width=.69\textwidth]{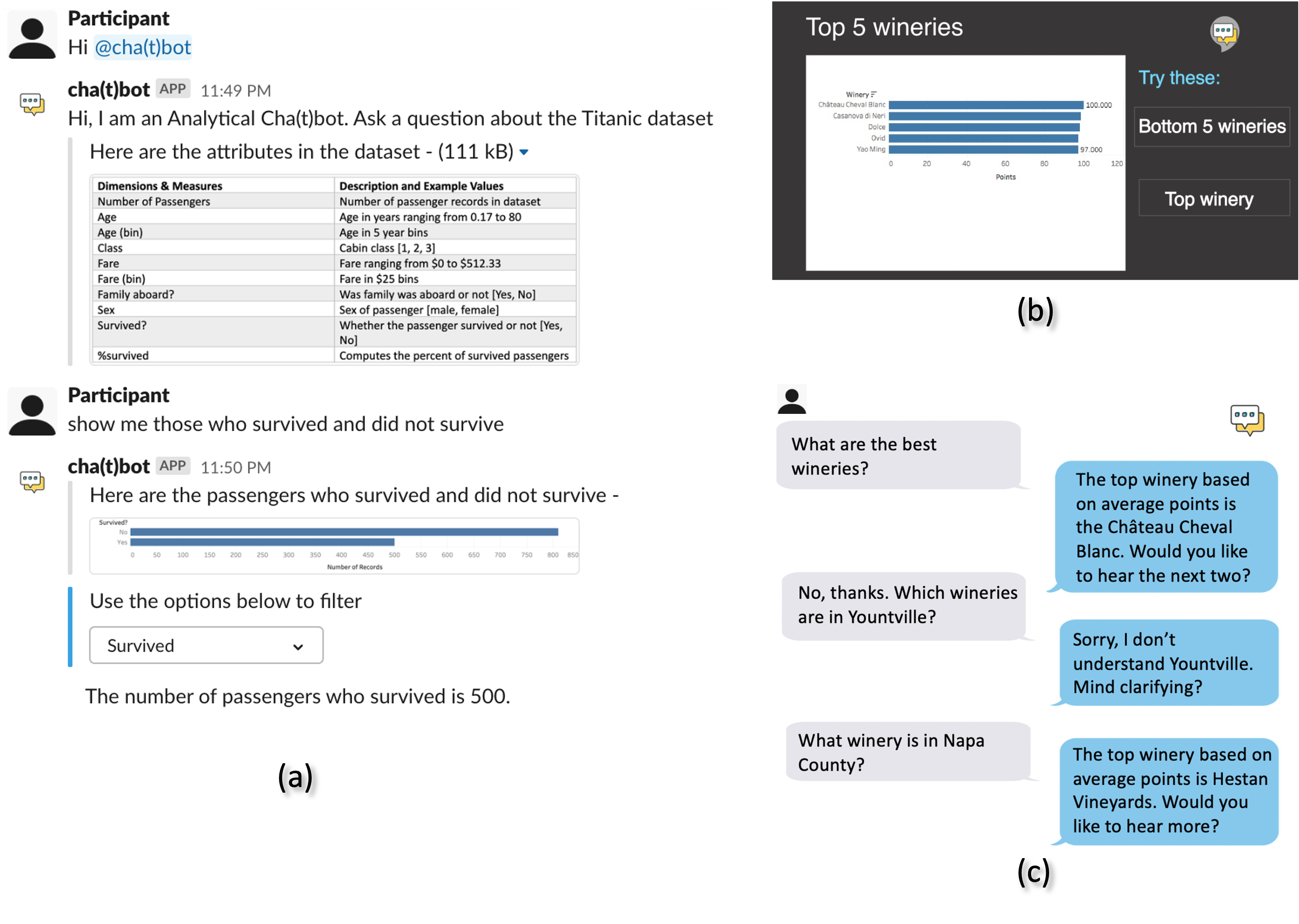}
  \caption{Participants conversing with various analytical chatbot prototypes. (a) A Slack chatbot showing an interactive message with a drop-down menu to help a user refine a previous response within the conversation thread. (b) An Echo Show chatbot simulator screen showing the top 5 wineries result along with two other follow-up utterance options on the right side of the screen. (c) Interaction with an Echo chatbot. The grey text bubbles indicate voice transcripts from the participants while the blue ones are from the chatbot. Follow-up questions and feedback from the chatbot encourage conversational behavior.}
  \Description[Participants conversing with various analytical chatbot prototypes.]{Participants conversing with various analytical chatbot prototypes. (a) A Slack chatbot showing an interactive message with a drop-down menu to help a user refine a previous response within the conversation thread. (b) An Echo Show chatbot simulator screen showing the top 5 wineries result along with two other follow-up utterance options on the right side of the screen. (c) Interaction with an Echo chatbot. The grey text bubbles indicate voice transcripts from the participants while the blue ones are from the chatbot. Follow-up questions and feedback from the chatbot encourage conversational behavior.}
  \label{fig:teaser}
 \end{teaserfigure}

\maketitle

\section{Introduction}
Conversational interfaces (CIs) such as smart assistants and chatbots have become prevalent for tasks ranging from simple fact-finding (e.g., asking for the weather) to question-and-answer scenarios such as making a restaurant reservation~\cite{atkinson:2010,sidnell2012handbook}. CIs constitute a distinctive form of interaction that borrows patterns from natural human conversation. With access to online resources, increased computational power, and machine-learning, CIs have come a long way from early natural language (NL) programs that were fraught with difficulty in user understanding~\cite{weizenbaum:1966}; they are now more conversational and understand reasonably complex utterances within known contexts~\cite{moorebook}.

Recently, natural language interfaces (NLIs) for visual analysis tools have garnered interest in supporting expressive ways for users to interact with their data and see results expressed as visualizations~\cite{setlur2016eviza,hoque2017applying,datatone,kumar2016towards,askdata,powerbi,thoughtspot,ibmwatson}. Users interact with a dataset or a visualization and can change the data display by filtering, navigating, and seeking details-on-demand. In these information-seeking conversations, the user may express their intent using NL input, and the system provides visualization responses. The analytical experience focuses on keeping the user in the flow of conversation. These interfaces are often designed for a specific platform or modality, with user intent understanding constrained by the domain of the knowledge base or context in which the interaction occurs. Furthermore, these conversational interfaces tend to focus on NL only as an input mechanism, not as part of the system response.

The promise that NL will make visual analysis tools more approachable has led to a proliferation of new potential entry points, platforms, and styles of interaction. One emerging interaction modality is the \emph{analytical chatbot}, a software application that engages in a back and forth NL dialogue with the user about data~\cite{hoon2020interfacing, fast2018iris, zhi2020gamebot, bieliauskas2017conversational, kassel2018valletto}. Like other types of chatbots, analytical chatbots are designed to simulate the way a human would act as a conversational partner, and therefore need to employ NL as both an input and output mechanism. They may additionally employ visualizations in their responses. When compared to existing NLIs for visual analysis, analytical chatbots have a different style of interaction and more ``agent-like'' behavior.

The emergence of analytical bots as mediators of data analysis activities presents new challenges and opportunities, some of which we investigate in this work. Merely repurposing how user intent is interpreted for one type of NLI in another does not always lead to precise interpretation. Additionally, we need to consider the interplay of NL and visualization components in how a bot responds to user questions. To build functionally intuitive analytical chatbots, we need to understand how users interact in these environments and develop design principles that can guide appropriate system responses in relation to utterance intent. \change{While there are general design guidelines for chatbot interfaces, in this paper, we wanted to explore how users interact with \textit{analytical} chatbot systems through natural language, and how modality affects both user interaction and behavior.}

Chatbot design often draws inspiration from human-to-human conversation and mechanisms that facilitate the exchange of information between speaker and listener. In such conversations, there is an expectation that the information shared is relevant and that intentions are conveyed. Grice’s Cooperative Principle (CP)~\cite{Grice1975-GRILA-2} states that participants in a conversation normally attempt to be truthful, relevant, concise, and clear. Consider this conversation snippet:

\begin{quote}
Lizzie: Is there another carton of juice?\\
Milo: I’m going to the supermarket in a few minutes!
\end{quote}

A human who reads the above conversation will easily infer that at the moment, there is no juice and that juice will be bought from the supermarket soon. Examples like these prompted Grice to propose various maxims where the CP explains the implication process. Grice argued that the generation and perception of implicatures are based on the following principle: ``Make your conversational contribution such as is required, at the stage at which it occurs, by the accepted purpose or direction of the talk exchange in which you are engaged.'' Though these Gricean Maxims have provided some guidance for human-computer mediated communication~\cite{herring2010}, little work has explored how to support cooperative conversation when a user is specifically exploring data with the help of an agent. In this cooperative framework, the question arises: when is it appropriate to introduce visualization versus language? When asking a question, we all are familiar with when an answer is too detailed or too terse. 
Because we are social beings with experience in conversation, we know what an appropriate response is and what the implications are when someone deviates from the norm. So how \emph{does} one converse with an analytical chatbot? What are the expectations of system behavior and interaction that support a cooperative conversation for data exploration? Are there differences in these user expectations across modalities and platforms?

\subsection{Research Questions}
Our primary goal is to explore how platform and modality differences influence users' conversational behaviors and system expectations when exploring and asking questions about data. Towards this goal, we ran a series of studies designed around best practices for both text- and voice-based CIs. We consider three platforms (voice-only, voice with visual responses, and text-based).
Specifically, our studies aim to address the following research questions:

\begin{itemize}
\item RQ1 - NL utterances: What are the characteristics of NL questions that users ask through text vs. voice?
What types of ambiguous and underspecified questions do they ask with these modalities?

\item RQ2 - Response expectations: What would users expect as a reasonable response? When do users expect only a text or voice response? When do they want charts to be shown along with a text or voice response? What are users' expectations of the charts shown in response to NL questions?

\item RQ3 - Modalities for repair: When the result is unexpected, how do users expect to repair the system behavior?
\end{itemize}

\subsection{Contributions}
This paper explores conversational patterns and expectations as users interact with analytical chatbots in various text- and voice-based platforms during data exploration. Specifically, the contributions of our paper are:

\begin{itemize}
\item  Revisiting Gricean Maxims, we explore design principles for supporting cooperative behavior and expectations in chatbot conversations that are specific to data exploration. 

\item We conducted a series of Wizard of Oz (WoZ) studies using three modalities: voice-only, voice with charts, and text with charts on Slack~\cite{slack} to better understand how users explore data through NL interaction. Findings from the studies show that analytical chatbot experiences constitute a distinctive set of user interaction behaviors and expectations. These observations provide additional context when employing Gricean Maxims as a guideline for conversational behavior during data exploration. 

\item  Based on observations from the WoZ studies, we identified and implemented a subset of design variants in three CI platforms -- Slack (text with charts), Echo Show (voice with charts), and Echo (voice-only).

\item We subsequently conducted an evaluation of these three prototypes to identify design implications and guidelines for creating useful experiences with analytical chatbots.
\end{itemize}

\section{Related Work}
We explore related work on NLIs for visual analysis and more specifically, analytical chatbots.

\subsection{NLIs for Visual Analysis}
NLIs have recently become popular as a means of interaction with data and may offer a lower barrier to entry compared to other interaction modalities. These conversational analytics tools automatically produce or modify visualizations in response to NL questions about data. DataTone~\cite{datatone} introduced ambiguity widgets, allowing users a means of repair when the system makes incorrect responses to ambiguous input. Eviza~\cite{setlur2016eviza} and Evizeon~\cite{hoque2017applying} supported ongoing analytical conversation by enabling follow-on queries via language pragmatics. Orko~\cite{srinivasan2017orko} extended these concepts to voice interaction and network diagrams, and InChorus~\cite{srinivasan2020inchorus} developed a framework for multimodal interactions involving both touch and voice. Additional systems that employ NL interaction with visualization include Articulate~\cite{kumar2016towards}, Analyza~\cite{dhamdhere2017}, and Text-to-Viz~\cite{cui2019text}.
Many conversational interaction concepts have also been deployed in commerical visualization tools (e.g., \cite{ibmwatson},  \cite{powerbi}, \cite{askdata}, \cite{thoughtspot}). All of these systems focus on NL as an input mechanism, where the system output is one or more charts. While many of the learnings from these systems may apply to chatbot interfaces, chatbots have a different interaction style and are expected to hold a natural language dialogue with the user.

Research has also investigated how natural language could be used to describe visualizations and data facts, potentially informing the design of an analytical chatbot's language responses. Srinivasan et al.~\cite{srinivasan2018augmenting} illustrated how visualizations augmented with interactive NL data facts could support exploratory analysis. Similarly, Wang et al.~\cite{wang2019datashot} generated automatic fact sheets containing both visualizations and NL, and Liu et al.~\cite{liu2020autocaption} generated automatic chart captions by employing a deep learning algorithm and NL templates. Longer narratives to express causality relationships were explored by Choudhry et al.~\cite{choudhry2020once}. Studies by Lima and Barbosa~\cite{lima2020vismaker} suggest that organizing visualization recommendations by the NL questions that they answer may help users understand recommendation content. Furthermore, empirical work on how people describe data insights and visualizations (e.g. Henkin and Turkay's~\cite{henkin2020words} research on scatterplot verbalizations) can serve as a foundation for automatic approaches to natural language generation. 

These conversational analytics and recommendation systems demonstrate value for NL as both an input and output modality for interaction with analytical tools. However, none of them specifically explore a chatbot style of interaction.


\subsection{Analytical Chatbots}
Chatbots have become a popular means of interactions in many applications, with some of the earliest ones being rule-based~\cite{weizenbaum:1966} and recent ones employing learning-based approaches~\cite{bordes:2016,Dodge2016EvaluatingPQ,Kannan2016SmartRA,li-etal-2016-deep,serban:2016,vinyals:2015}. 
\change{
For factors known to influence the user experience of chatbots, the reader is referred to several recent surveys~\cite{chaves2021should, rapp2021human, mygland2021affordances, dobrowsky2021influence}. For example, Rapp et al. reported that realistic user expectations, relevance and timeliness of chatbot responses, and the chatbot's personality, transparency, and social interaction style all influence human trust. Similarly, Chaves and Gerosa~\cite{chaves2021should} describe how human-like social characteristics such as conversational intelligence and manners may benefit the user experience. However, human-like characteristics are perceived more favorably only up to a point; chatbots with imperfect human-like behaviors may trigger an uncanny valley effect~\cite{dobrowsky2021influence, ciechanowski2019shades}. Text and voice interaction modalities are particularly relevant to our work. A comparative study of voice and text-based interaction with chatbots~\cite{rzepka2021voice} found that voice was generally preferred in terms of cognitive effort, enjoyment, efficiency, and satisfaction, but this was influenced by goal-directedness of the task.
}

\change{Most closely related to our work are analytical chatbots for answering questions with data}. Hoon et al.'s~\cite{hoon2020interfacing} `analytics bot' augmented a data dashboard so that users could ask additional questions about the data, but the chatbot produced only text responses, not visualizations. 
Visual Dialog~\cite{Das_2017_CVPR} was an AI agent that could hold a dialog between a computer and a human, discussing visual content. The characteristics of the conversation included temporal continuity and grounding the visual content in the conversational exchange. A two-person chat data-collection protocol was used to curate a large-scale dataset (VisDial) containing question-answer pairs and to train a set of neural encoders to create a visual chatbot application. Our paper explores a similar goal of enabling conversational interaction, including visual artifacts, but in our case, the focus is to support answering questions about data.

In the data space, Fast et al.~\cite{fast2018iris} introduced a chatbot for Data Science with a limited ability to plot statistical charts and Valetto~\cite{kassel2018valletto} introduced an analytical chatbot for tablets, employing a chat-style interface side by side with a chart.
GameBot~\cite{zhi2020gamebot}, a chatbot for sports data, demonstrated how narrative text and visualizations could be integrated in chatbot responses. 
Bieliauskas and Schreiber~\cite{bieliauskas2017conversational} illustrated how an analytical chatbot could be integrated into team messaging environments such as Slack. Their chatbot could adjust filters and metrics in a network visualization juxtaposed next to the chat window. Both of these latter chatbots were domain-specific (sports or software engineering) and their utility was not evaluated.

Most similar to our work are studies investigating user expectations around analytical chatbots. Kassel and Rohs~\cite{kassel2019talk} explored expectations around chatbot responses with the Valetto prototype~\cite{kassel2018valletto}, introducing an `answer space' framework varying across level of statistical detail and whether the answers were descriptive or explanatory. They found that people's statistical knowledge influenced the style of answers they preferred and that it was important to match the level of detail in the chatbot's answer to the user's language. Hearst and Tory~\cite{hearst2019would} conducted a series of crowd-sourced studies to understand when users expected text versus chart responses to predefined data questions. They found a split in people's preferences, with approximately 40\% preferring not to see charts in their analytical chatbot conversations. Those who did appreciate charts generally preferred to see more data than they specifically requested to provide context. In a similar experiment, Hearst et al.~\cite{hearst2019toward} explored how analytics systems should respond to natural language queries with vague terms like `high' or `expensive.' \change{Zhi~\cite{zhi2020coupling} compared usability of three response formats in an interactive chatbot: text only, text with visualizations, and text with interactive visualizations. Results showed a strong preference for interactive visualizations that enable access to more information than requested.}

Our research employs a series of exploratory Wizard Of Oz and prototype evaluation studies to investigate people's expectations around chatbot interaction. Like Kassel and Rohs~\cite{kassel2019talk}, \change{we found that the level of detail in the chatbot response influences user assessments of appropriateness. Mirroring Hearst and Tory~\cite{hearst2019would} and Zhi~\cite{zhi2020coupling}, our results show that users tend to prefer interactive visualizations, and value context and additional information in chatbot answers.} We extend this line of research beyond level of detail and types of context, to consider both text and voice input and output modalities, use of message threading, and the interplay between text and visualization responses.

\section{Analytical Chatbot Design Principles}
The goal of our work is to understand how we can support users' data exploration in chatbot interfaces for commonly available modalities, ranging from text interaction with visual responses in a medium like Slack~\cite{slack} to voice-based interaction commonly found in smart assistants~\cite{alexa,googlenest}. 

Understanding the structure of a single utterance and its semantic content is not enough to have a complete understanding of the conversational context. Pragmatic reasoning that understands the context and intent of the conversation lends itself to a more engaging experience~\cite{ChakrabartiLuger2015}. The interaction design space for implementing conversational experiences for chatbots can be vast and vague. Despite the importance of pragmatic processing, evaluating the quality of conversation is difficult to determine. While grammars and well-defined language rules can address syntactic and semantic handling of individual input utterances, there is no gold standard to evaluate the quality of a chatbot with respect to its conversational behavior. In order to ground the possible variants in this conversational design space to specific conversational characteristics, \change{we employ Grice's cooperative principles}~\cite{Grice1975-GRILA-2}. \change{The principles describe how speakers act cooperatively to be mutually understood for effective communication.} Grice divided the cooperative principle into four conversational maxims.
We describe each of the maxims and how we apply them to chatbot design, specifically guidelines for effective system responses and interaction behavior.

\begin{itemize}
\item \textbf{Maxim of Quantity}: Be informative. Provide all the information necessary for the purpose of the current conversational exchange. Do not make your contribution more informative than is required, but ensure that the response addresses the intent in the question. For example, the conversation snippet below has just the right amount of information about the nearest store along with its opening time. 

\vspace{2mm}
\noindent \scstart{human} \textit{``When does the nearest grocery store open?''}
\newline \noindent\scstart{chatbot} \textit{``The nearest grocery store is at 48 Main Street and it opens at 8:00 am.''} \vspace{2mm}

Violations of this maxim are either a terse chatbot response saying, ``8:00 am'' or too detailed a response such as, ``There are three grocery stores located within a radius of 10 miles. The nearest store is 1.4 miles away at 48 Main Street and opens at 8:00 am.''
\vspace{2mm}

\item \textbf{Maxims of Quality}: Be truthful. Avoid stating information that you believe might be wrong, unless there is some compelling reason to do so. If you do choose to include it, then provide a disclaimer that points your doubts regarding this information. Avoid including information that cannot be supported by evidence. For example, in the conversation snippet below, the chatbot greets the human and sets the appropriate expectations regarding its capabilities of understanding the conversation.

\vspace{2mm}
\noindent \scstart{chatbot} \textit{``Welcome! I'm a virtual assistant that can help you book a concert ticket. You can ask me simple questions or follow my lead. Remember that I'm not a human and can't understand everything. Shall we start?''}
\newline \noindent\scstart{human} \textit{``Sure!''} 
\vspace{2mm}

A violation of this maxim is a chatbot greeting that simply says, ``Hi! You can ask me anything about the concert.''  This example does not set up the conversation for success as the chatbot is not transparent about its capabilities, leading to unrealistic user expectations.
\vspace{2mm}

\item \textbf{Maxim of Relation}: Be relevant. Make sure that all the information you provide is relevant to the current exchange and omit irrelevant information. For example, in the conversation snippet below, even though the human did not respond to the chatbot's initial question, the chatbot provides a response relevant to the human's question. Providing a follow-up inquiry after the relevant response is a useful way of directing the human back to the original question that the chatbot posed or indicating the presence of other related tasks.

\vspace{2mm}
\noindent \scstart{chatbot} \textit{``Would you like to book an appointment?''}
\newline \noindent\scstart{human} \textit{``When's the next availability?''} 
\newline \noindent \scstart{chatbot} \textit{``The next available appointment is at 11 am on Friday. Would you like to make an appointment or modify an existing one?''}
\vspace{2mm}

A violation of this maxim is a chatbot response, ``Please answer yes or no'' to the human's question, ``When's the next availability?'' In this case, the chatbot is not providing a relevant response to the human and continues to focus on its original intent of booking an appointment.

\vspace{2mm}

\item  \textbf{Maxims of Manner}: Be clear and concise. Avoid obscurity of expression and ambiguous language that is difficult to understand. Ask for clarification or follow-up inquiry to support conversation turns. Unlike the previous three maxims that primarily focus on what is said during the conversational exchange, the Maxim of Manner focuses on \emph{how} that exchange occurs. For example, in the conversation snippet below, the chatbot is conveying its thought process to the human clearly by sharing and requesting for information in a turn-by-turn manner.

\vspace{2mm}
\noindent \scstart{chatbot} \textit{``Please hold while I connect you to a representative.''}
\newline \noindent (After 20 seconds)
\newline \noindent\scstart{chatbot} \textit{``Sorry, no one's available right now. Would you like me to send an email? They will respond in 24 hours.''} 
\newline \noindent \scstart{human} \textit{``Yes!''}
\newline \noindent \scstart{chatbot} \textit{``Great. To send the email, I first need some information about you. What's your first name?''}
\vspace{2mm}

A violation of this maxim is a chatbot response that simply ends the conversation without providing a follow-up option, for example, ``Sorry, no one's available right now. Bye-bye!''
\end{itemize}

For the purpose of analytical chatbot design,  Gricean Maxims provide a basic framework for determining the various components of a conversation. We draw inspiration from an established set of best practices for identifying and implementing cooperative chatbot behaviors~\cite{Habermas1984-HABTTO,saygin:2002,cassell:2005,baptiste:2019}. We identify the following conversational design patterns (DP) with their relevant maxims:

 \begin{itemize}
 \item \textbf{DP1: Greeting and orientation:} When the user first interacts with the chatbot, the greeting needs to clearly convey what purpose the chatbot serves (Maxims of Manner and Quantity). 
 
 \item \textbf{DP2: Turn-taking:} Conversations should be a back and forth exchange so that users do not need to specify all the details at once~\cite{moore2018}. The chatbot should avoid dead-end responses and provide prompts to move the conversation forward. It should understand context between sequential utterances and anaphoric references to prior utterances (e.g., ``What did you mean by that?'', ``how about adding coffee beans to the order?'') (Maxim of Manner).

\item \textbf{DP3: Acknowledgements and confirmations:} To build trust, acknowledgments need to be provided as feedback indicating that the user's input was received. The chatbot should ask the user to repeat the query or clarify the system response in situations when the chatbot’s confidence in recognizing the intent is low (Maxims of Quality and Relation).

\item \textbf{DP4: Concise and relevant responses:} To minimize cognitive effort, chatbot responses should be concise and to the point based on the user's intent. Lengthy content can be broken into chunks with the most relevant chunk returned first. Users should be able to add follow-up clarification or request more information, for example, by clicking on a button or asking an explicit follow-up query (Maxims of Quantity and Manner).
 \end{itemize}

We acknowledge that while Gricean Maxims help frame expectations for chatbot design, there are some criticisms of the theory. For instance, the Gricean Maxims do not specifically provide guidance for handling conversational ambiguity (i.e., queries with more than one possible interpretation) or misinterpretation. These cases of failure in conversational implicature may be due to linguistic parsing issues, failure to understand the user's actual intent, or simply misunderstanding of idioms of the language. The only general guidance that Gricean Maxims provide is to have the user and/or the chatbot restate or clarify the question~\cite{hadi:2013}. However, in the NLI space, there is a precedence in how visual analysis tools handle underspecification (i.e., queries with missing information such as an attribute name, date value or analytical operation) and ambiguity. Some systems interpret user intent through simple pragmatics in analytical interaction using contextual inferencing, wherein the context established by the preceding dialog is used to create a complete utterance, in combination with information from the data domain~\cite{datatone,askdata,setlur2016eviza,hoque2017applying,srinivasan2017orko}. Most NLI tools provide targeted textual feedback with the system responses, along with ambiguity widgets that enable the user to both repair and refine the system choices. We hence include two additional design patterns that are specific to \emph{analytical} conversation within the chatbot interaction space:

\begin{itemize}
\item \textbf{DP5: Ambiguous and underspecified utterance handling:}  When chatbots encounter an ambiguous or underspecified utterance, they need to provide feedback to the user explaining their interpretation of the utterance and how it was handled. For data exploration, ambiguous utterances can arise when there are multiple ways of interpreting the intent~\cite{setlur2016eviza}. Underspecified utterances have missing information that needs to be filled to create a valid query that can be executed against the underlying datasource to generate a system response~\cite{setlur2019inferencing}. For example, for the query, ``which products are doing \emph{well}?'', the word `well' is both underspecified and ambiguous as the user did not mention which data attribute(s) to associate it with and what data range of values to filter the query to. In this case, the chatbot could infer \texttt{Sales} and/or \texttt{Profit} as the relevant attributes with some pre-defined range filters. The chatbot should present a concise text or verbal explanation of its inferences that is relevant to the context of the data. If there are other viable interpretations, the chatbot should provide follow-up options to present alternatives to the user. If disambiguation is not possible, the chatbot should request help from the user to explicitly clarify the utterance. A message introducing the clarification request could include phrases such as, ``Did you mean...'', ``Was this answer helpful?'', or ``This is what I could find...''

\item \textbf{DP6: Refinement and repair:} Complementary to the handling of ambiguity and underspecification, chatbots should provide interface affordances (visual or language) so users can refine and repair system choices and interpretations. In a GUI context, graphical elements, such as buttons, images, and menus, could be mixed into the interaction alongside NL input~\cite{schegloff:1977}. These elements can enable the user to choose alternative analytical functions (e.g., `average' instead of `count'), options to change or include other data attributes, and value filters for updating the system response and visualization. Voice-only chatbots need to elicit clarification through a series of verbal actions that are presented one at a time. For example, ``how about adjusting \emph{young} to be 12 and under instead?''
\end{itemize}

\section{Study 1: Evaluating Interaction Behavior}
\begin{figure}
  \centering   
    \includegraphics[width=\linewidth]{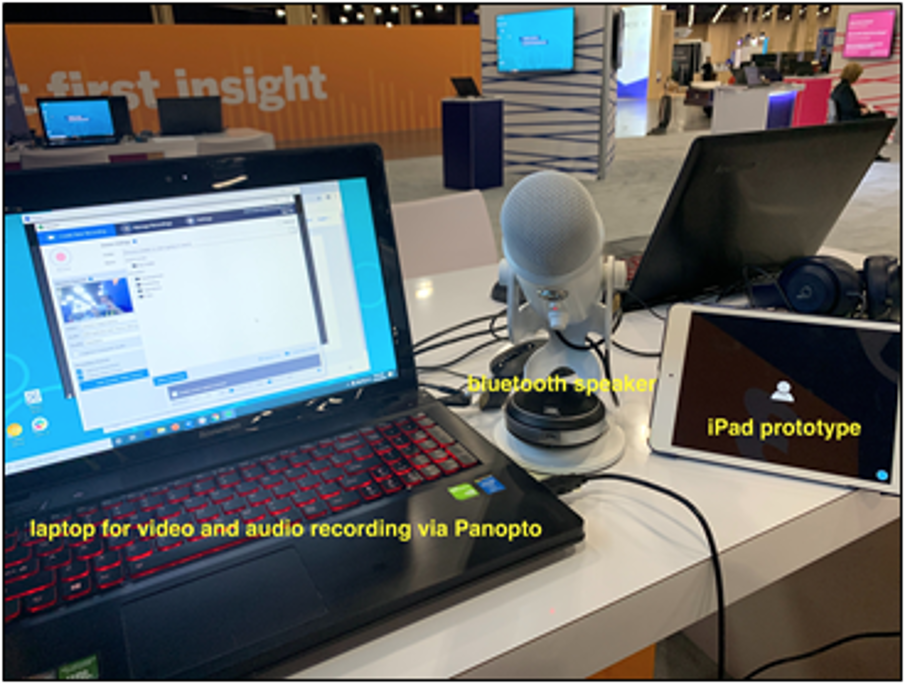} 
    \caption{An example setup of the iPad variant in the studies}
    \Description[An example setup of the iPad variant]{An example setup of the iPad variant in the studies}
    \label{fig:setup} 
\end{figure}

To fully explore the expressibility of queries and responses, we ran the studies as Wizard of Oz simulations, where two human wizards produced visualizations and responses to the participants' input. We used a dual-wizard protocol to reduce difficulty of the wizard role. One wizard operated Tableau to generate visualizations, and the 2nd wizard provided text or voice responses based on a template of responses (\autoref{fig:template}), with the complete version in the supplementary material.
Below is the setup information for each study. An example is shown in \autoref{fig:setup}.

We conducted three exploratory Wizard of Oz studies to observe how people use NL interaction for visual analysis on communication platforms such as Slack and smart assistant devices such as Alexa. We collected NL utterances, plus qualitative data on user expectations. Each study investigated a different modality - (Study 1a) text interaction using Slack, (Study 1b) voice interaction using a Bluetooth speaker device, and (Study 1c) voice interaction using an iPad. Although the studies were conducted separately, we present them together as the method, task, and setup was largely the same. Any differences are called out in the sections below.

\subsection{Participants}
A total of $30$ volunteer participants ($18$ female, $12$ male) took part in the studies, and none of them participated more than once. All participants were fluent in English. The participants had a variety of job backgrounds with visual analytics experience - administrator, supply chain consultant, legal, user researcher, engineering leader, data analyst, senior manager of BI, product manager, technical program manager, and a marketing manager. The participants signed up at an industry tech conference or were recruited from a local town email group, with the criteria being that they were conversant in English and were familiar with using any chatbot or smart assistant device. Participation in the study was voluntary, and participants were \change{offered a conference tote bag and a water bottle for their time}.
We use the notation [P\#] when referring to participants in these studies.

\subsection{Prototypes}\label{sec:prototype}

\subsubsection{\change{Application of Design Patterns}}

\change{We first summarize how we apply the six design patterns to the study variants with additional details based on the different modalities described in more detail in each of the three study sections.}

\begin{itemize}
\item \change{\textbf{DP1: Greeting and orientation:} To address the Maxims of Manner and Quantity, participants are greeted in voice and / or text with a metadata summary of the data source they can ask questions about.}
\vspace{2mm}
\item \change{\textbf{DP2: Turn-taking:} To address the Maxim of Manner, we employ threading in Slack and pose follow-up questions through voice to encourage turn-taking.}
\vspace{2mm}
\item \change{\textbf{DP3: Acknowledgements and confirmations:} To address the Maxims of Quality and Relation, we rely on a template of text and verbal acknowledgements and confirmations that are consistent with each study type for various analytical expressions.}
\vspace{2mm} 
\item \change{\textbf{DP4: Concise and relevant responses:} To address the Maxims of Quantity and Manner, we rely on a template of text and verbal responses that are crafted to be relevant to the questions. To stay concise, fact-finding questions are answered with a single text response without the display of a chart, or with a verbal response.}

\item \change{\textbf{DP5: Ambiguous \& underspecified utterance handling:} For handling ambiguity and underspecification, we include responses that attempt to clarify the wizard's interpretation with additional text or verbal explanation and a prompt for the participant to clarify.}
 \vspace{2mm}
\item \change{\textbf{DP6: Refinement and repair:} Participants were provided the option to re-clarify their questions or amend the wizard's response by typing or asking a follow-up question.}

\end{itemize}

\paragraph{\textbf{(Study 1a) Text interaction using Slack}} The participant and the wizard each had a Mac laptop with a Slack app connected to the same workspace. The participant was shown a welcome message and a description of the data source (\textbf{DP1}). They also had access to a laminated information sheet about the datasource. The participant interacted with the data by typing a question into Slack. The questions could be of aggregation, group, filter, limit, and sort expression types as found in Tableau. The wizard responded by typing a response based on a pre-defined template of responses for each corresponding expression type (\textbf{DP3}). The wizard then pasted an image of the corresponding visualization generated via Tableau for that question (using the Mojave OS Screenshot app on the Mac) into the Slack channel. Note that single answer responses in Tableau were just pasted as text into Slack (without any chart response) (\textbf{DP4}).

Slack has additional features that help with conversational interaction (\textbf{DP2}). The first is message threading that facilitates focused follow-up conversations inside a `flex pane' next to the main chat pane~\cite{slackthreads}. Threads help to organize information by making the public channels more readable and moving discussions about discrete topics into their own workspace (\textbf{DP4}). 

\paragraph{\textbf{(Study 1b) Voice interaction using a Bluetooth speaker}} The wizard used a laptop connected to a Bluetooth speaker and Amazon Polly~\cite{polly} to convert the text response into computer-generated speech output. The Bluetooth speaker welcomed the participant with a brief summary of the data source (\textbf{DP1}). They also had access to a laminated information sheet about the data source. The participant initiated the system by prefacing the start of the conversation with ``Hey <chatbot greeting> (anonymized)'' so that the wizard could distinguish between general chatter and questions intended to be parsed by the chatbot. The participant interacted with the data by verbally asking a question about the data. The questions could be of aggregation, group, filter, limit, and sort expression types as found in Tableau. The wizard responded by typing a response into Polly based on a pre-defined template of responses for each corresponding expression type (\textbf{DP3, DP4}). Responses were played on the Bluetooth speaker as audio output to the participant. Upon completion of a task, the wizard added a follow-up question like, ``Is there anything else I can help you with?'' to support conversational turns (\textbf{DP2}).

\paragraph{\textbf{(Study 1c) Voice interaction using an iPad + Bluetooth speaker setup}} The wizard used a Mac laptop connected to an iPad via Bluetooth. A separate Bluetooth speaker provided audio output, while the iPad functioned as a display to show visualization responses. The wizard used Amazon Polly to convert the text response into computer-generated speech output. The iPad welcomed the participant with a brief summary of the data source shown on the screen (\textbf{DP1}). The participant also had access to a laminated information sheet about the datasource. The participant initiated the system by saying ``Hey <chatbot greeting> (anonymized)'' so that the wizard could distinguish between general chatter and questions intended to be parsed by the chatbot. They interacted with the data by verbally asking a question about the data. The questions could be of aggregation, group, filter, limit, and sort expression types as found in Tableau. The wizard responded by typing a response into Polly based on a pre-defined template of responses for each corresponding expression type (\textbf{DP3, DP4}). The wizard then took a screenshot of the corresponding visualization generated via Tableau using the Screenshot app on the Mac. The wizard sent the chart image to the iPad via the Message app on the Mac laptop. Note that single-answer responses in Tableau were just sent as verbal responses without an accompanying chart image. Similar to Study 1b, upon completion of a task, the wizard added a follow-up question to support conversational turns (\textbf{DP2}).

\begin{figure}
  \centering   
    \includegraphics[width=\linewidth]{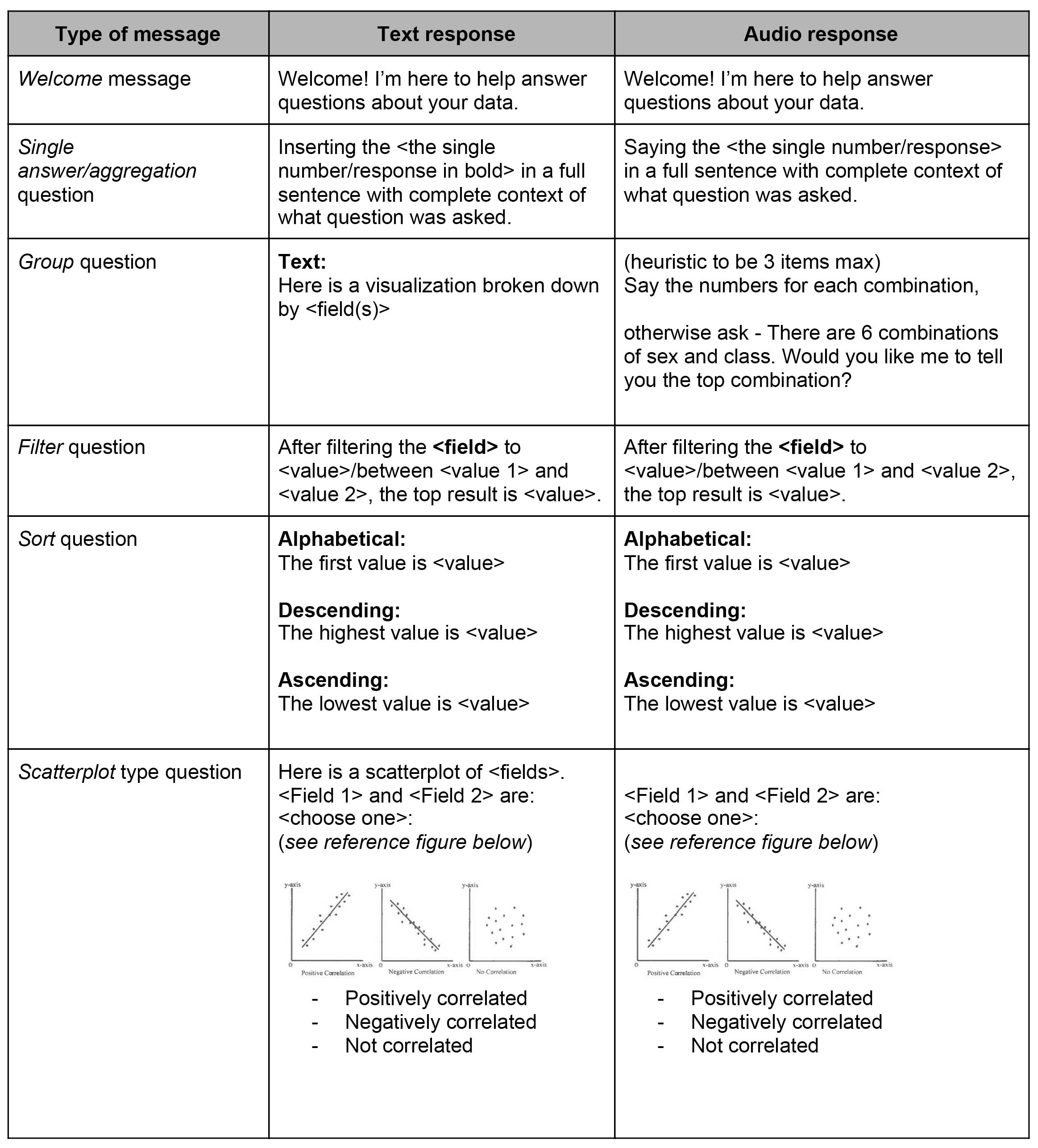} 
    \caption{A subset of template responses used by the wizard}
    \Description[A subset of template responses]{A subset of template responses used by the wizard}
    \label{fig:template} 
\end{figure}

\subsection{Task and Data}
Participants were asked to explore a dataset about passengers onboard the Titanic ship. They were asked to focus on questions containing attributes from the dataset, including passenger age, fare, class information, and Boolean attributes to indicate whether a passenger survived or not, and had family aboard or not. 

\begin{figure*}[t]
\centering 
    \subfigure[Study 1a - Slack]{\fbox{\includegraphics[width=0.32\linewidth]{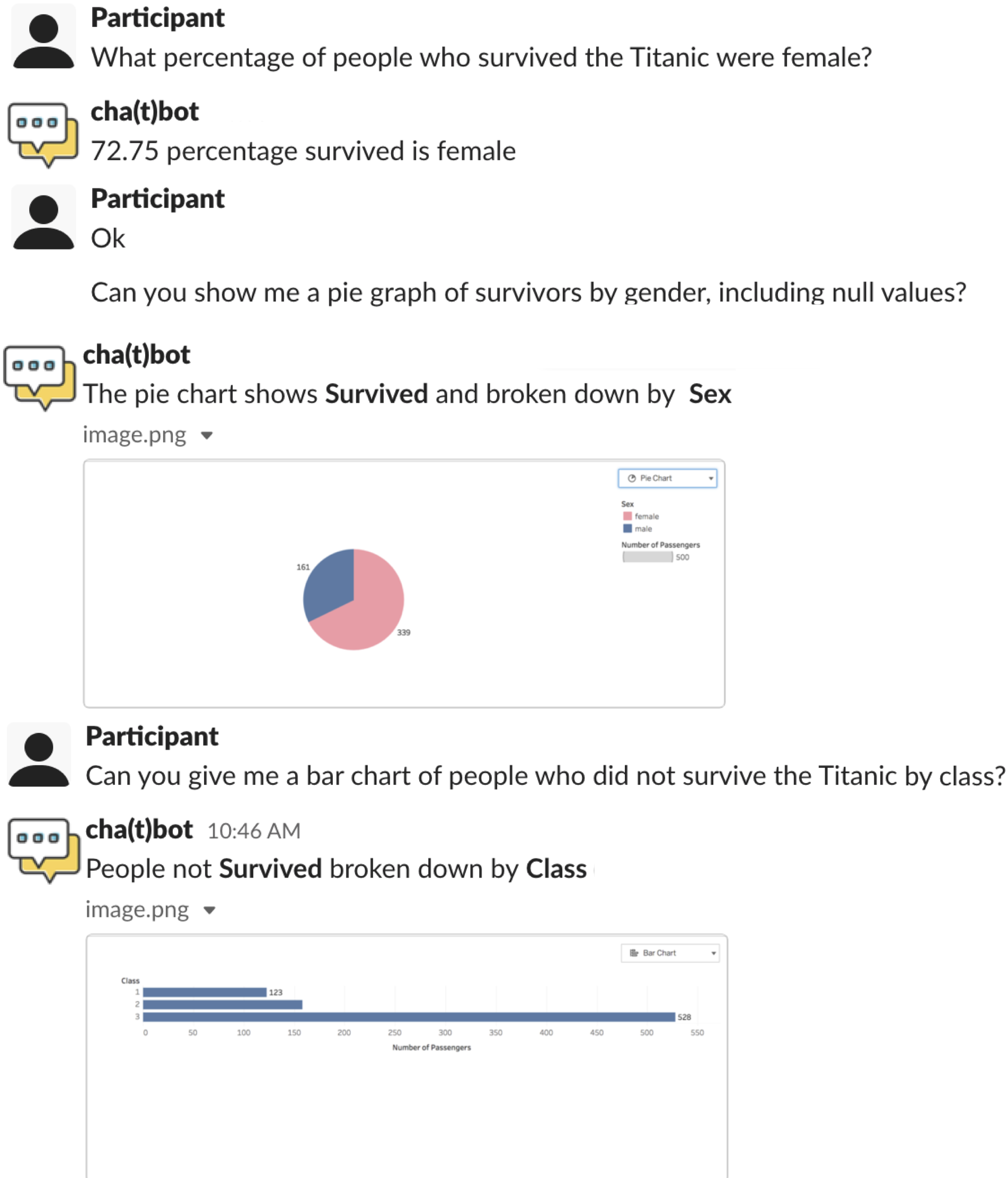}}}
        \hspace{1mm}
   \subfigure[Study 1b - Voice]{\fbox{ \includegraphics[width=0.3\linewidth]{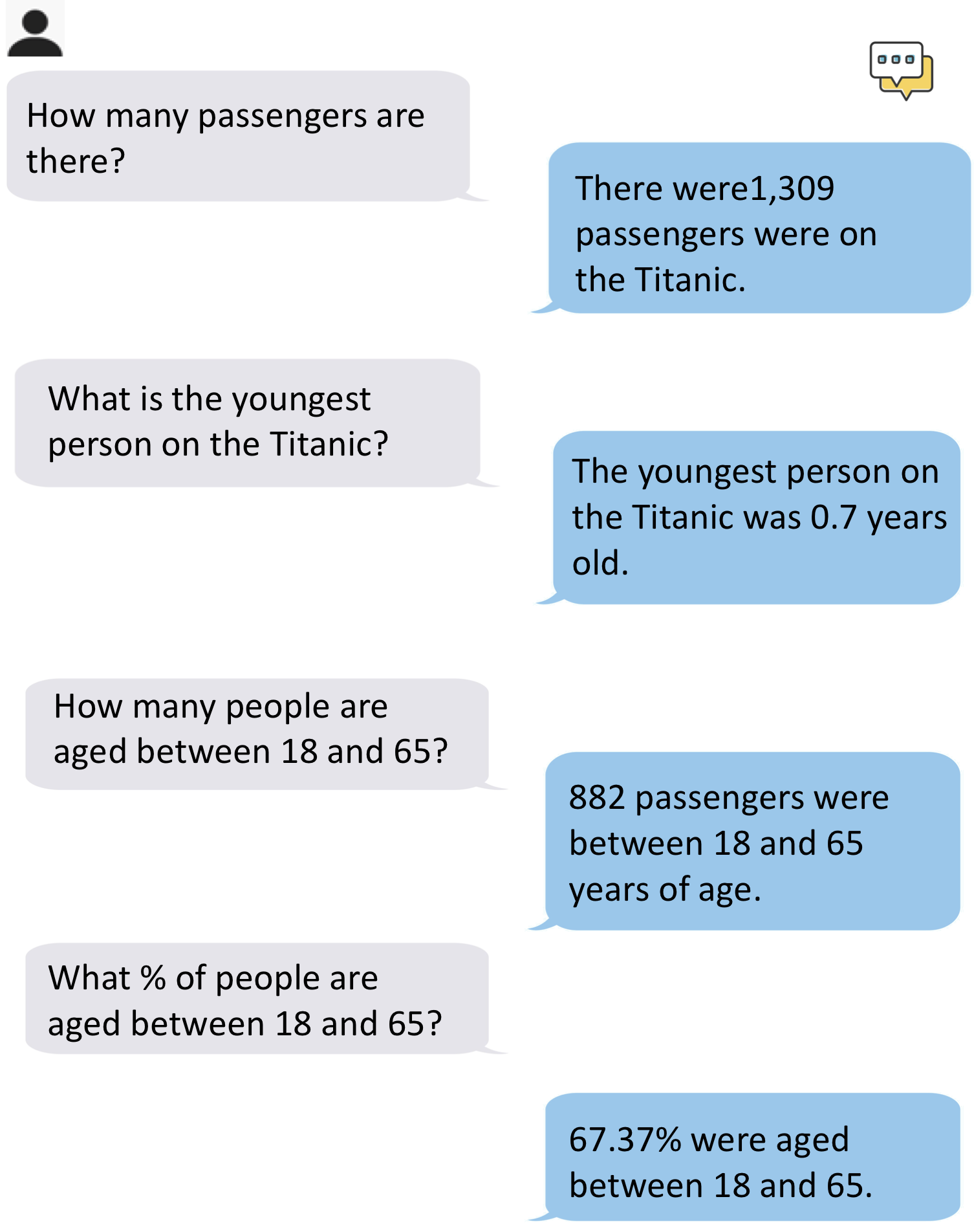}}}
    \hspace{1mm}
      \subfigure[Study 1c - Voice with iPad]{\fbox{\includegraphics[width=0.29\linewidth]{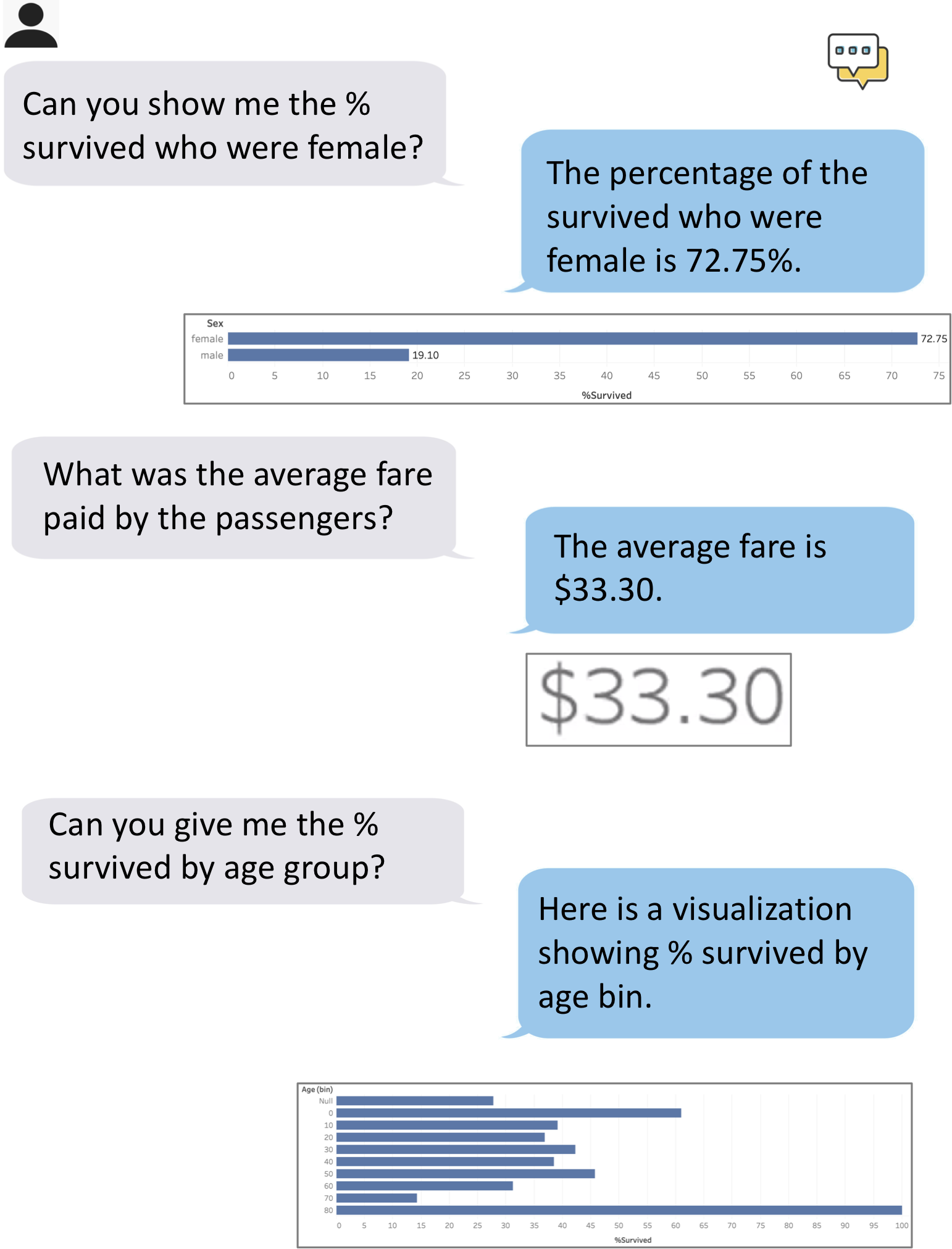}}}
            \caption{Conversation snippets from Study 1. (b) and (c): The grey text bubbles indicate voice transcripts from the participants while the blue ones are from the chatbot. Visualizations are displayed alongside the system responses in (a) and (c).}
              \Description[Conversation snippets from Study 1]{Conversation snippets from Study 1. (b) and (c): The grey text bubbles indicate voice transcripts from the participants while the blue ones are from the chatbot. Visualizations are displayed alongside the system responses in (a) and (c).}
    \label{fig:study1examples}
\end{figure*}

\subsection{Procedure}
We conducted $10$ sessions in each study, each lasting 25 minutes. Four staff members supported each session: one facilitator, one note taker, and two wizards. Participants were not made aware of the Wizard prior to participation. The facilitator followed a script. Participants were first introduced to the study and we asked about their background and role. They were then given instructions and spent most of the session interacting with the system by entering text questions in Slack or asking voice-based questions, and then observing the resulting visualizations plus text or audio responses. 

We employed a question-asking protocol to elicit qualitative feedback. While the system was ``thinking,'' the facilitator asked the participants what they expected as a response to their input, and then when the response arrived, the facilitator asked for the participant's feedback. Given that the responses were manually generated by the wizard, there was no built-in logic for ambiguous and underspecified utterance handling or repair. Instead, participants were asked to restate a modified follow-up utterance if the response was not what they expected (\textbf{DP5, DP6}). Participants were told at the end of the session that the system was a simulation. We then wrapped up the session during the last 5-10 minutes, getting their overall feedback about the prototype.

\subsection{Data collection and analysis}
Natural language utterances were collected with audio recordings of the voice input and Slack history for the text input. Sessions were screen-recorded and audio-recorded. A notetaker was present in most sessions to take field notes. Field notes were expanded to a video log after the study through partial transcription of the videos. The video log (and raw video for reference) was then qualitatively coded to look for themes and trends.

\section{Study 1 Findings}

\begin{figure*}
    \includegraphics[width=0.8\linewidth]{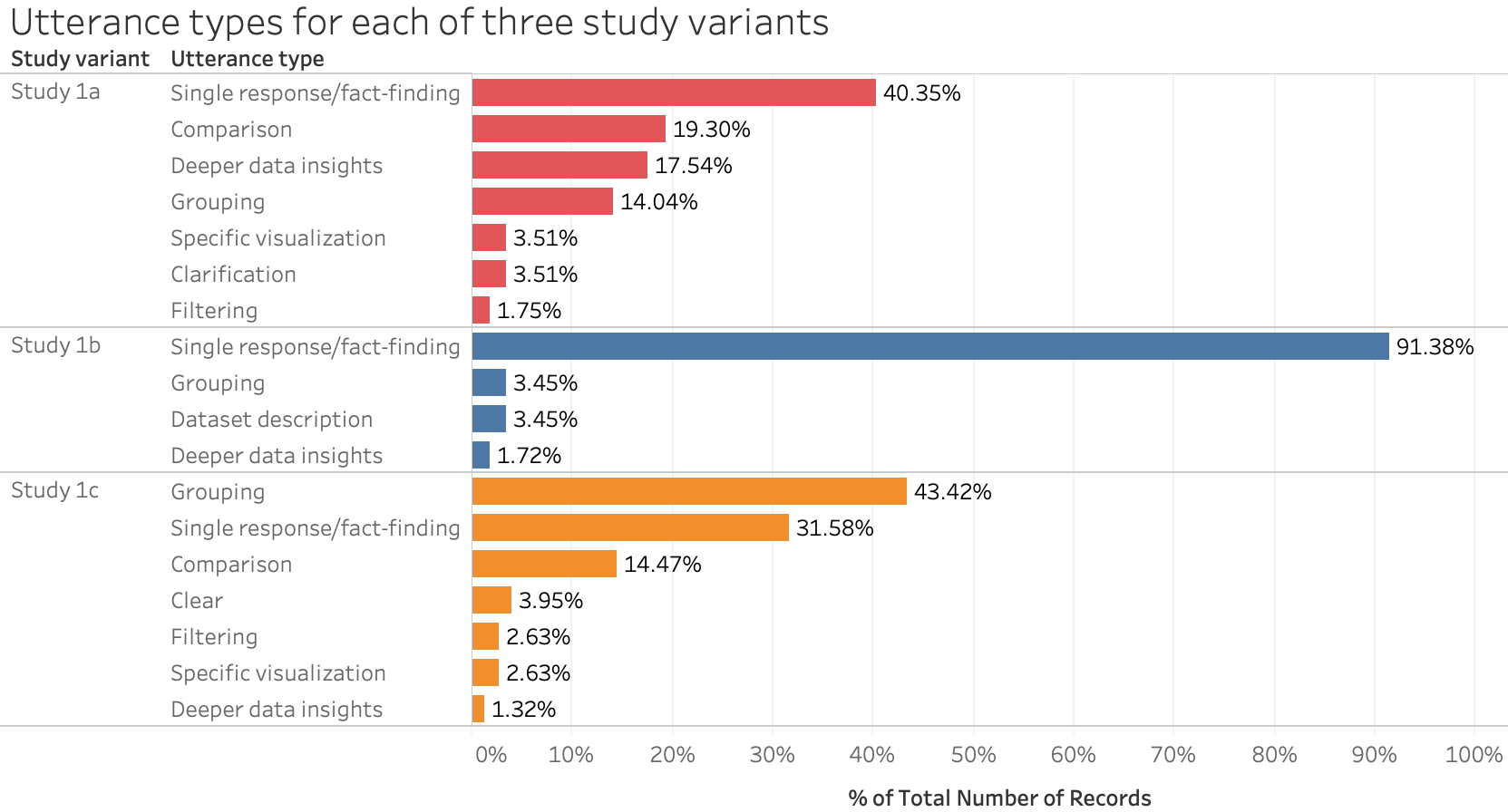} 
    \includegraphics[width=0.8\linewidth]{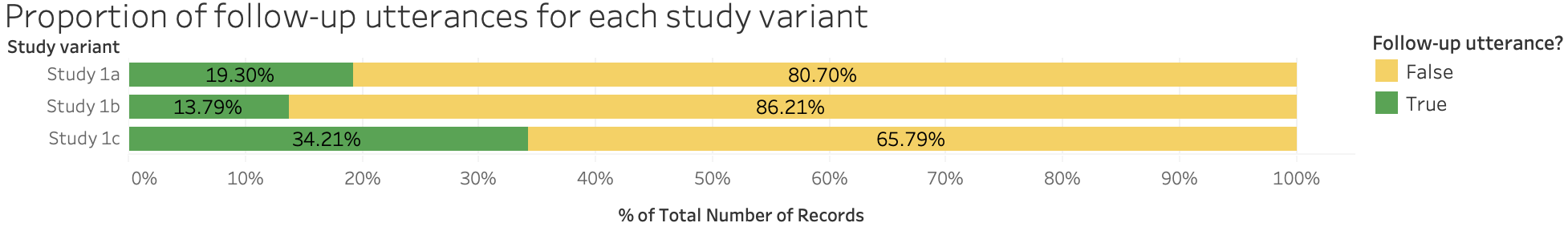} 
    \caption{Utterance classification from Studies 1a-c. Top: Voice modalities elicited a greater proportion of fact-finding questions, especially in Study 1b. The analytical categories expressed were varied with the need for deeper analytical insights in Study 1a. Bottom: In general, there were fewer follow-up utterances across all the three studies, with Study 1b (voice-only) having the least proportion.}
    
      \Description[Utterance classification and follow-up utterances from Studies 1a-c.]{Utterance classification from Studies 1a-c. Top: Voice modalities elicited a greater proportion of fact-finding questions, especially in Study 1b. The analytical categories expressed were varied with the need for deeper analytical insights in Study 1a. Bottom: In general, there were fewer follow-up utterances across all the three studies, with Study 1b (voice-only) having the least proportion.}
    \label{fig:analysis}
\end{figure*}

For each study, we categorized the input utterances based on the type of analytical intent they referred to. The categories included the five basic database operations found in VizQL~\cite{vizql} along with other intents such as `clear' for starting a new conversation, `compare' for comparing two values in a field, `clarification' for wanting to clarify the system's response, and asking for a specific visualization type. The full set of classification types is available in the supplementary material. Examples of conversation snippets from the studies are shown in \autoref{fig:study1examples}. We also classified whether the utterances were follow-up utterances to a previous conversation thread or not. These data differed in interesting ways for the three variants, as shown in \autoref{fig:analysis} and summarized in the following sections.

\subsection{(Study 1a) Text interaction using Slack}
Ten participants asked the Slack prototype 124 utterances in total (Avg 10.6 utterances per session). Based on coding of the videos and the notes, 40.4\% of the utterances were manually classified as fact-finding, expecting a single response such as a number or ``yes/no'' (e.g., P15 - ``how many families survived in total?''). 19.3\% of the utterances were that of a comparison intent where participants wanted to compare a set of values for a given attribute (e.g., P18 - ``Can you show me a chart of survival \% compared to age?'' ). A small proportion (14.4\%) of these utterances involved grouping by an attribute (e.g., ``what was the average age for the female survivors?''). Interestingly, there were several examples (17.5\%) where the participant wanted deeper insights about the data (e.g., P15 - ``have there been outliers per class?'', P16 - ``How much more likely was a passenger to survive if they were in first class?'').

19.3\% of the initial utterances had follow-up utterances. Several follow-ups involved reformulating the original utterance when the system response was unexpected. For example, P15 reformulated the utterance ``can you show me this graph in clusters?'' with a follow-up, ``can you show me this graph in bins?''. P14 reformulated the original utterance ``can I see all fares paid for men?'' with a follow-up, ``all fares paid by men?'' when they found the Gantt chart returned by the first utterance to be unacceptable. They hoped that simplifying the utterance would result in something more favorable (which did not happen). Others used follow-up utterances as a means to help clarify what they were seeing. For example, P18 asked ``if you were female, are you more likely to have survived?'' with a follow-up ``Why?''.

Text interaction via Slack elicited a variety of analytical questions beyond simple fact-finding, often involving multi-turn conversation threads. This led us to further investigate this modality in a later study (\autoref{sec:study2}).

\subsection{(Study 1b) Voice interaction using a Bluetooth speaker}
Ten participants asked the voice-only prototype 103 utterances in total (Avg 9.72 utterances per session). Based on coding of the videos and the notes, a majority (91.4\%) of the utterances were manually classified as fact-finding, expecting a single response such as a number or ``yes/no'' (e.g., ``did any passengers survive on Titanic?''). This was much higher than in the iPad and Slack studies, suggesting that a voice-only chatbot would be used primarily for lookup tasks rather than deeper analysis. A small number of utterances involved grouping by an attribute (e.g., ``what is the distribution of age bin per passenger?'') and asking for more information about the Titanic dataset (e.g., ``what is this dataset?''), with 3.5\% for each. Interestingly, there was one fact-finding question that expected the system to provide deeper analytical insights, asked by P10 - ``Is there any outlier for fare in class 1?''

The voice-only study had a low incidence of follow-up utterances, amounting to 13.8\%. Among those follow-up utterances, a majority of them were also fact-finding in nature (e.g., ``What is the age of these paying 0 in class 1?''), with a few utterances requesting a number to be swapped out for a percentage (e.g., ``how many of the passengers that survived paid more than \$300?'', followed by ``what's the percentage?''). One participant (P12) tested the prototype with some trick questions just to assess the limitations of the system by asking questions such as ``did anyone have the name almo?'' and ``what about aria?'' even though they were told that the dataset did not contain the names of passengers.

\subsection{(Study 1c) Voice interaction with an iPad + Bluetooth speaker}
Ten participants asked the prototype 110 utterances in total (Avg 10.08 utterances per session). Based on coding of the videos and the notes, utterances were manually classified into one of the following categories: 43.4\% Grouping by an attribute (e.g., ``Show survival rate by fare class''), 31.6\% Fact-finding, expecting a single response such as a number or ``yes/no'' (e.g., ``How many female passengers survived?''), 14.5\% Comparison across values for an attribute (e.g., ``\% of men and women who survived''), with a smaller percentage of the remaining utterances either being resetting the context of the conversation, explicitly requesting a chart type (e.g., ``Box plot for the fare''), or asking a deeper insight or reasoning (e.g., ``What's the key factors that indicate somebody survived or not'').

In the iPad study, 34.2\% of the utterances were classified as follow-ups. 22.4\% of those follow-up utterances involved adding an attribute to the current visualization (e.g., ``can you split it by number of people survived?''). A small number of follow-up utterances involved swapping out an attribute with another (e.g., ``Switch class by fare'') and filtering out nulls (e.g.,``Remove null from this dataset''). Interestingly, a new type of follow-up utterance was also observed where a user asked a follow-up fact-finding question about the visualization (e.g., ``Average fare these women paid'').

\subsection{User expectations}
\label{sec:expectations}
Based on participants' alouds, we observed some common user expectations spanning across the chatbot variants:
\noindent\textbf{Automatically filter nulls}: Several participants across the Slack and iPad variants expected the system to automatically filter out nulls, with accompanying language indicating that the filter was applied to the visualization response. 

\noindent\textbf{Provide context to fact-finding questions}: In the iPad variant, there were several utterances for which the system behavior was not satisfactory. P02 asked, ``was the first class more expensive than others?'' Upon seeing the response, they said, ``A complicated Gantt chart with no explanation wasn't that helpful.'' When asked if a simple yes/no response would have been preferred, they replied that the Boolean response would probably be more useful than the Gantt chart, but would still expect some additional context. As another example, for an utterance ``what \% of passengers in cabin class 1 survived?'' a response ``62\% of class 1 survived when compared to 43\% in class 2 and 26\% in class 3'' is more useful than just ``62\%.'' In the voice-only variant, participants were expecting the system to parrot back some version of the question, especially those questions that could be answered by a single number or a yes/no response; here the context confirms that the system correctly understood the user's request.

\noindent\textbf{Support query expressibility}: One of the challenges while designing a natural language interface is the high variability in how people express questions. While we saw follow-up threads in the Slack and iPad variants, the utterances in the voice-only variant were found to be precise, self-contained fact-finding questions such as ``how many people who were 50 or older were on the titanic?'' As P04 said - ``It is an interesting concept, can see non tech-savvy people use this [...] with voice that people frame their questions linguistically in a certain way. I'd be concerned in both text and voice, but with voice, there are more nuances. I'd be concerned whether the responses would be the same if asked differently.''

\noindent\textbf{Semantics is important}: Participants used a variety of synonyms and related concepts in their utterances. For example, P06 asked in the iPad variant, ``How many families are fully lost on the boat,'' where ``fully lost'' pertained to ``not survived.'' P4 asked ``Average fare these women paid,'' where paid refers to ``Fare.'' Recognizing synonyms and concepts would help enhance the recognizability of these types of utterances, in addition to providing self-service tools for people to add domain-specific concepts with their datasets.

\noindent\textbf{Support repair and refinement}: Many of the follow-up utterances for the Slack and iPad variants involved adding an additional attribute to the analysis, swapping out a number for a percentage to do a comparison or filter out information. Even in the voice-only variant, follow-up questions often involved a fact-finding inquiry based on the current context. When designing natural language systems with these various voice/text/visual modalities, it is important to set the right expectations to the user that follow-up utterances and clarification are supported. This aligns with existing design guidelines suggested for chatbots, including the one on Alexa, where the suggestion is to design responses with intents followed with a question such as ``Do you want to know more?'' or ``Is there anything else I can help you with?''.

\noindent\textbf{Understand follow-up utterances vs. resetting the context}: Very few people used terms such as ``clear'' and ``start over'' to explicitly reset the context, even though that information was part of the instructions. Several participants used anaphora such as ``that chart'' to refer to the current context. This pattern of behavior was more pronounced in the Slack and iPad variants. We could leverage Slack threads to explicitly provide feedback to the system that a user intends to follow-up on a previous conversation. However, the problem of automatically detecting follow-up vs. a new utterance is more challenging in voice-based interaction as the system would need to reliably detect anaphora.

\noindent\textbf{Support interactive visualizations}: A few participants expressed that they would have liked the visualizations to be interactive or editable via an authoring tool. P14 said, ``Some interactivity might help to reformulate the results.'' The screenshots we used (from Ask Data~\cite{askdata}) showed a drop-down widget for choosing a different visualization type that was not clickable and set false expectations about interactivity. P18 said, ``if it (the visualization) is static, I wouldn't expect there to be a drop-down box.''

\noindent\textbf{Provide a text description accompanying the visualization responses}: Several participants did not notice that the prototype was speaking to them. They often forgot what it said, and when looking at the visualization, they forgot what they were looking at. Participants wanted a text description, feedback pills, or a caption describing the visualization. This information could also show the attributes the system used to generate the visualization, helping users to determine whether the system correctly interpreted their question.

\noindent\textbf{Enable deeper insights and reasoning}: With the chart variants, especially Slack, participants (P15, P16, P18) asked several ``why'' questions about observations such as outliers and trends. Extending the capabilities of analytical conversation interfaces to not only provide the ``what'', but the ``why'' and ``how'' from the data, could help facilitate richer and deeper analytical workflows with such tools.

\noindent\textbf{Integrate chatbots into other visual analysis workflows}: Chatbots are conducive for question-answering, but participants also expected them to integrate into other aspects of the analytical workflow, such as creating dashboards and saving the results to a workbook. P03 said in their exit interview, ``Could we throw vizzes to dashboard too while we are asking questions [...] so this tool can be used as the dashboard builder. Clients who don't have IT departmental infrastructure can use this tool for automating some of that stuff.  We use a lot of auditing and can use this tool there.''

\section{Study 2: Evaluating Analytical Chatbot Interfaces} \label{sec:study2}
Based on our observations of user behaviors and preferences from the three sets of Wizard of Oz studies, we implemented three working analytical chatbot systems on Slack (supporting text and images), the Echo Show (supporting voice and images), and the Echo (supporting voice only) platforms. We ran an exploratory study with each of these platforms to collect qualitative data on how users interact with these chatbots and the types of data exploration behaviors they exhibit. Unlike the Wizard of Oz studies in Study 1, where a human wizard controlled the interaction behavior with the participant, Study 2 implemented three working chatbot systems that automated the system responses.

\begin{figure}
  \centering
  \includegraphics[width=\linewidth]{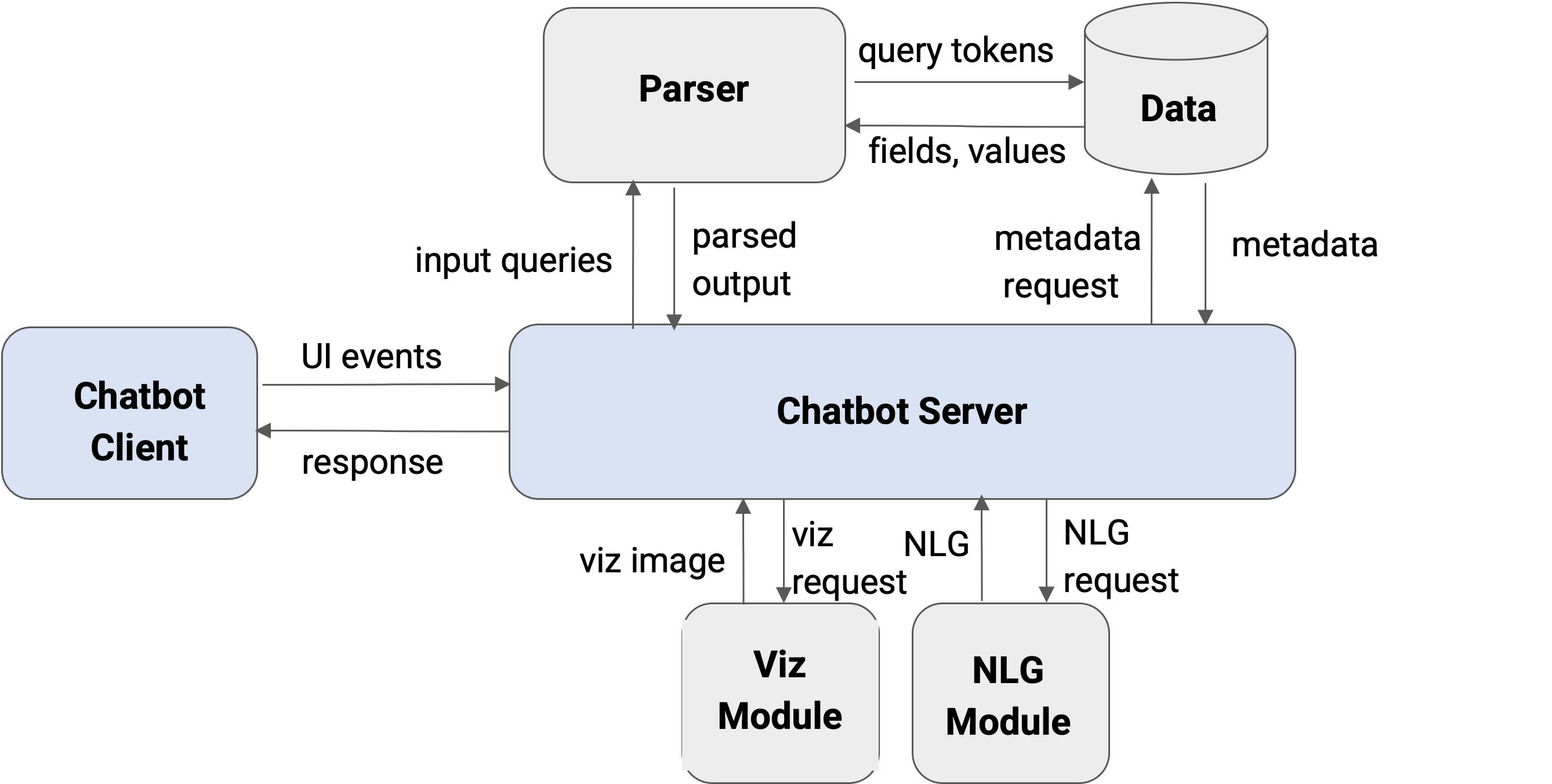}
  \caption{System overview of the chatbot architecture}
  \Description[System overview of the chatbot architecture]{System overview of the chatbot architecture}
  \label{fig:system}
\end{figure}

\subsection{Method}
We chose a between-subjects design to avoid learning and fatigue effects. Participants were randomly assigned to a chatbot condition, and either the Titanic passenger dataset or a wines review dataset~\cite{kagglewines}. The task, procedure, data collection, and analysis were similar to those in Study 1 with differences documented below. 

 We conducted 10 sessions per condition, each lasting 25 minutes. Two staff members supported each session: one facilitator and one notetaker. The facilitator followed the same experiment script from Study 1. Participants were first introduced to the study and we asked about their background and role. They were then given instructions and spent most of the session interacting with the system and observing the resulting visualizations and text responses. We employed the same question-asking protocol from Study 1 to elicit qualitative feedback. We then wrapped up the session during the last 5-10 minutes getting their overall feedback about the prototype.

All sessions were screen-recorded and audio-recorded. For Slack, NL utterances were collected from the conversation history logs.  Field notes were expanded to a video log after the study through partial transcription of the videos. The video log (and raw video for reference) was then qualitatively coded to look for themes and trends. All studies took place outdoors with masks on to conform with COVID-19 social distancing protocol. 

\subsubsection{Participants}
Ten participants took part in each of the three study variants, with a total of $30$ ($15$ female, $15$ male). Note that these participants were different from those who participated in Study 1. All participants were fluent in English and familiar with using a chatbot platform. Similar to the previous studies, the participants had a variety of job backgrounds with visual analytics experience ranging from a school staff member, graduate students, entrepreneurs, program managers, software engineers, and data analysts. Participants were recruited via a public mailing list for a local town. Participation in the study was voluntary and \change{were offered gourmet cupcakes from a local bakery for their time}. We use the notation [P'\#.Condition], where `Condition' is ``Slack,'' ``Echo Show,'' or ``Echo'' to contextualize quotes with the condition the participant experienced.

\subsection{System Implementation}

The chatbot systems employ a node.js~\cite{nodejs} client-server architecture and have the following general components (\autoref{fig:system}):
\begin{itemize}
\item \textbf{Chatbot Client:} Listens to user greetings, interaction events and message events from the Chatbot Server (\textbf{DP1}). In the case of Slack and Echo Show platforms, the interface also displays native interactive widgets for surfacing ambiguity. 
\vspace{1mm}
\item \textbf{Chatbot Server:} The main application-specific server bridge between the Chatbot Client and the other components of the application. The server translates input client events (e.g., slack messages or voice commands) into appropriate API requests and responses into a format appropriate for the client. 
\vspace{1mm}
\item \textbf{Parser:} Parses input NL queries (text- and voice-based) into tokens based on an underlying grammar as implemented in Eviza~\cite{setlur2016eviza}. These tokens are resolved as data attributes and values (with information from the data source), or intent lexicons such as `trend' and `correlation' as well as modifiers such as `young' and `best'~\cite{Setlur2020SentifiersIV}. The parser also supports intent handling and infers underspecified or ambiguous information, similar to work in ~\cite{setlur2019inferencing}  (\textbf{DP5}). The server passes the parsed tokens to the Chatbot Server, so that the information can be used to generate a system response.
\vspace{1mm}
\item \textbf{Viz Module:} Generates images of data visualization results based on information such as chart type, intent strategy, data attributes, and values using Vegalite~\cite{vegalite} commands. This module is relevant to GUI-based chatbots such as Slack and the Echo Show.
\vspace{1mm}
\item \textbf{Natural Language Generation (NLG) Module:} Employs simple language templates for NLG with pre-defined placeholders to insert information for generating text- and voice-based system responses (\textbf{DP3}). Given that the application domain for these chatbot interactions uses a set of known analytical intents along with attributes and values from the underlying data, the space of linguistic variations is relatively small and the outputs can be specified using templates~\cite{reiter:2010}. We define the templates by referring to utterances from Study 1, along with utterances commonly supported across existing NLIs~\cite{setlur2016eviza,hoque2017applying,yu2019flowsense,setlur2019inferencing,narechania2020nl4dv} and sample utterances collected through studies investigating the use of NL to create or interact with data visualizations~\cite{tory2019mean,srinivasan2021collecting}. The grammar rules from the parser modules are used to aid in the NLG process, which involves ordering constituents of the NLG output and generating the right morphological forms (including verb conjugations and agreement)~\cite{reiter:1997}. 

\end{itemize}

\begin{figure}
  \centering   
    \includegraphics[width=\linewidth]{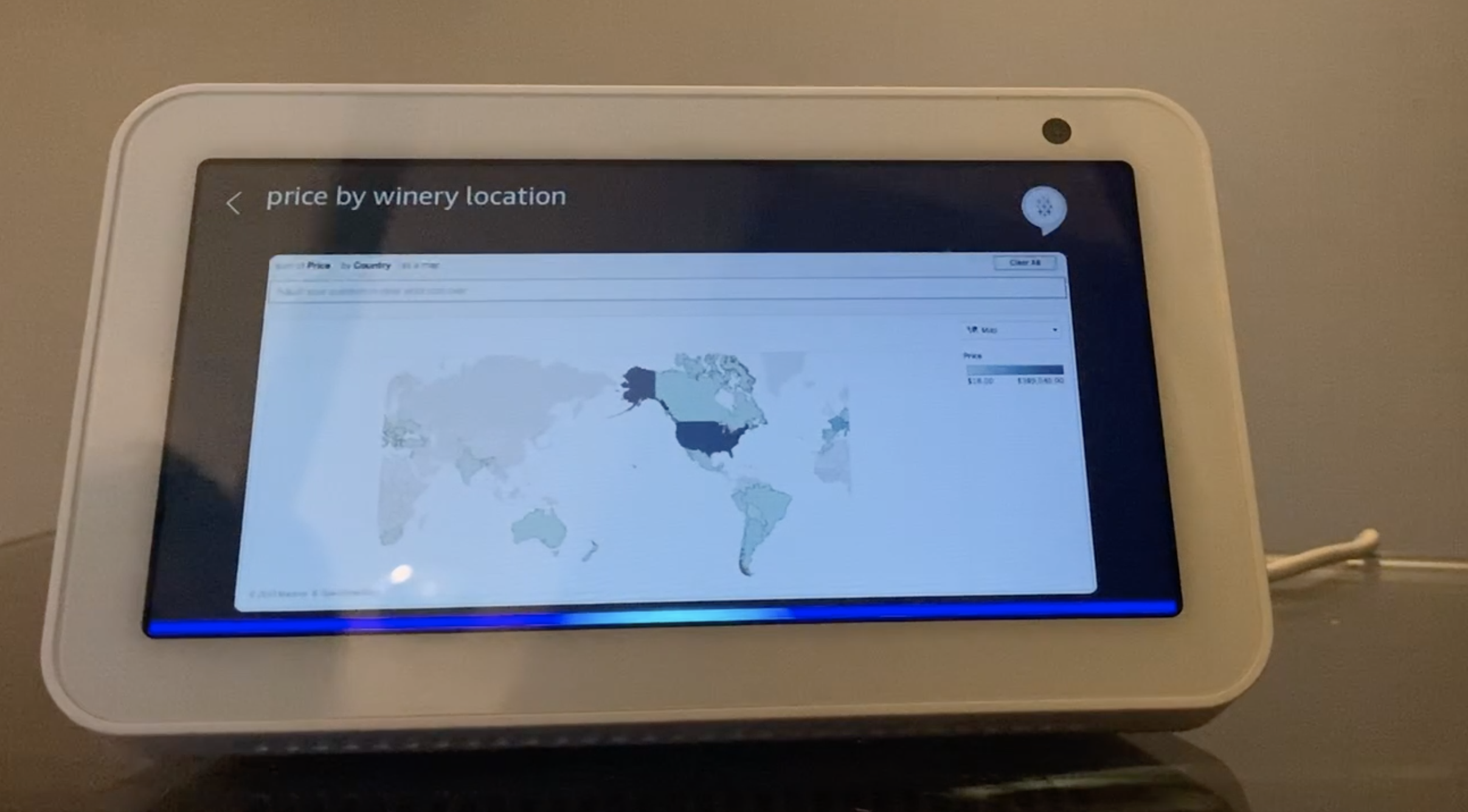} 
    \caption{Echo Show prototype showing a result for ``price by winery location'' and actively listening for the next utterance}
    \Description[Echo Show prototype]{Echo Show prototype showing a result for ``price by winery location'' and actively listening for the next utterance}
    \label{fig:study2echoshow} 
\end{figure}

The Slack chatbot uses the Slack API~\cite{slackapi} for listening to Slack events. Slack responses from the template used in Study 1 (\autoref{sec:prototype}) are passed as input to the prototype as a JSON file. The prototype automatically generates a system response as a new thread to the original top-level utterance when it detects follow-up questions (\textbf{DP2}); for example, when the user refers to the context in the previous utterance using anaphoric references such as ``that viz" or ``how about showing the response for first class instead.'' We did not provide any specific instructions to the participants about when to interact in threads since we wanted to observe their behavior without any priming. When a participant chose to respond in a thread, the Slackbot also automatically responded in the same thread. When the participant decided to type a question in the main channel, a new thread was automatically created with the corresponding system response (\textbf{DP3, DP4}).

The prototype utilizes Slack's interactive messaging framework~\cite{slackmessaging} that augments messages with interactive interface affordances such as buttons, menus, and custom actions for displaying ambiguity widgets (\textbf{DP6}), as seen in \autoref{fig:teaser}a. We implement two types of interactive widgets to accompany the chatbot responses: (1) a drop-down menu for filtering to specific values on the data domain; (2) a yes/no button option to clarify whether the response is expected when the input utterance is ambiguous (\textbf{DP5}). 

\begin{figure*}
    \includegraphics[width=\linewidth]{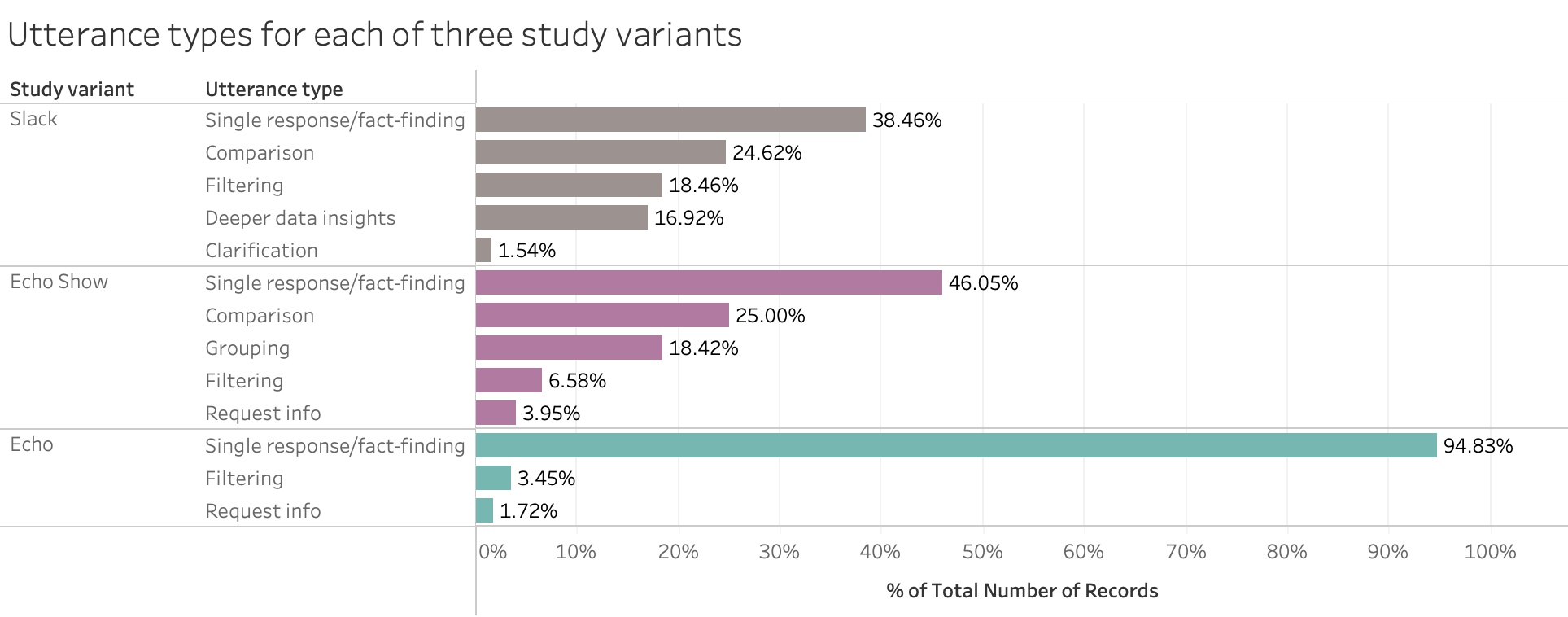} 
    \includegraphics[width=\linewidth]{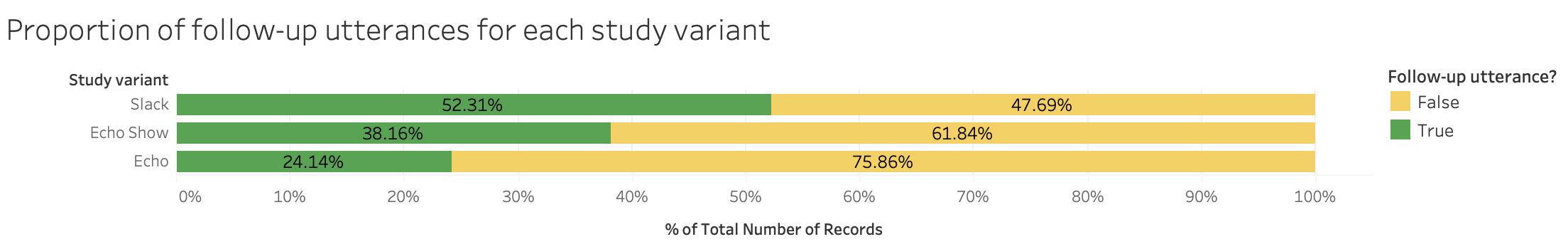} 
    \caption{\change{Utterance classification from the Slack, Echo Show, and Echo prototype studies. Top: Similar to findings in Study 1, voice modalities elicited a greater proportion of fact-finding questions, especially in the Echo chatbot. Bottom: Proportion of follow-up utterances across all three studies.}}
    \Description[Utterance classification from the Slack, Echo Show, and Echo prototype studies]{Utterance classification from the Slack, Echo Show, and Echo prototype studies. Top: Similar to findings in Study 1, voice modalities elicited a greater proportion of fact-finding questions, especially in the Echo chatbot. Bottom: Proportion of follow-up utterances across all three studies.}
    \label{fig:study2analysis}
\end{figure*}

The Echo Show and Echo chatbot systems have a similar implementation architecture to the Slack chatbot. However, rather than using a bespoke parser, the application employs the Alexa API~\cite{alexa} for parsing intents in the utterances. We activate a feature called \emph{Follow-Up Mode}~\cite{alexfollowup} that lets users make multiple requests, including follow-up inquiries without having to say the trigger phrase, ``hey, chatbot!'' each time a question is asked (\textbf{DP2}). Participants were instructed to use the trigger phase once at the beginning of the interaction session to set the Echo device in active listening mode, indicated by a blue halo light on the chatbot device (see \autoref{fig:study2echoshow}). Both the Echo Show and Echo chatbots provide verbal follow-up prompts to either continue or refine the current conversation, or ask a new question (\textbf{DP3, DP4}). The Echo Show can display a list of options on its touch screen based on  pre-defined display templates available for Alexa devices~\cite{alexadisplaytemplates} when it encounters ambiguous or underspecified utterances (\textbf{DP5, DP6}). We chose a popular US-English based female voice option called `Joanna'~\cite{alexavoice} for both the voice chatbots. 

\subsection{Study 2 Findings}
We summarize people's reactions to the three prototypes and examine the impact of their behavior as participants conversed. \autoref{fig:teaser} shows examples of participants conversing with the Slack, Echo Show, and Echo chatbots. \change{A summary of the utterance types and proportion of follow-up utterances is shown in \autoref{fig:study2analysis}.}

\subsubsection{Slack Study Findings}
In general, all participants were comfortable using Slack or a similar collaborative messaging platform. Many were curious what a real-time interaction with a chatbot would be like as several reported having used Slackbots that were either passive or served a one-time need like conducting a poll with a group. P'07.Slack remarked, ``I'm used to using a whole bunch of Slackbots like Polly (for polls) or seeing my Google calendar events for the day. This feels more interactive.'' 

Ten participants asked the Slack prototype 147 utterances in total (Avg 13.2 utterances per session). Similar to the procedure in Study 1, we manually classified the types of utterances based on coding of the videos and the notes. 38.46\% of the utterances were manually classified as fact-finding, expecting a single response such as a number or ``yes/no,'' 24.62\% of the utterances were that of a comparison intent where participants wanted to compare a set of values for a given attribute, 18.46\% of these utterances involved filtering by a data value, and 16.92\% involved a request for deeper insights about the data. The remaining small percentage of utterances were failure cases or a clarification.

\paragraph{\textbf{Effect of threads on conversational behavior}} 52.31\% of the utterances were follow-up conversations within the Slack threads. The average number of conversation turns\footnote{A type of organization in conversational discourse wherein the participant and the chatbot alternate in response to one another.} was $3.7$. People generally liked threaded responses for single word/number responses, but often found the real-estate in the Slack flex pane too small to read the visualizations. When the facilitator suggested that they could resize the pane to make it wider, the user experience improved considerably. P'03.Slack said, ``This is cool. The system responds in a thread and makes me want to ask something more about what I'm seeing here.'' P'07.Slack thought that threading helped them to easily refer back - ``I sometimes want to go back and see what the number was and I could search for my question and see the chat history around it.''  A few participants did not like the automatic threading and found it confusing. P'08.Slack commented, ``it's unclear where to place my comments as I need to think if I'm asking a follow-up or a new topic'' and  P'06.Slack said, ``Difficult to track new messages.'' On the other hand, participants found that the presence of widgets helped them focus their gaze to the bottom of the thread and see the responses in-situ while interacting with the widgets. P'04.Slack said, ``I liked being able to follow a discussion thread after interacting with the menu options. It helped keep the flow linear.''

\paragraph{\textbf{Utility of interactive widgets in the system responses}} $78.4\%$ of the total system responses contained one or more widgets and participants frequently interacted with them (76.8\% of total system responses containing widgets). Threading along with the widgets motivated participants to interact with the widgets as they preferred seeing the responses from those interactions in the thread. P'08.Slack said, ``I like to see the chatbot immediately respond in the thread when I play with the menu. That way, I can see all the results in one place for easy lookup.'' Generally, participants liked the drop-down menu to show alternative responses by filtering to other values. Having interactive widgets with threading prompted longer back-and-forth conversation, as P'09.Slack states, ``The menu made me want to click on it and when I saw the response in the thread, I wanted to choose other options from the drop-down. That made me want to ask something more about it.'' However, buttons to verify the expectations of the system responses got a mixed reaction. Some participants appreciated the affordance to provide feedback, ``I liked to click on \emph{Yes} to make sure that the chatbot remembers my preference. It knew what I meant when I asked for rich passengers.'' P'02.Slack exclaimed, ``Nice! I hit \emph{No}, and the system gives me a hint of how I can rephrase my question for better luck.'' Others found the buttons less useful; e.g., ``I don't need the buttons, but rather a prompt asking me if I want to rephrase and give me a hint directly. I do not want to click on the \emph{No} to get it [the chatbot] to follow up with me.'' [P'09.Slack]

\paragraph{\textbf{Types of utterances and failure cases}} Generally, we observed richer analytical conversations than just fact-finding or single-turn conversations. $39.5\%$ of the utterances were identified as being ambiguous across conditions. Participants often restated the utterance using a mixed-initiative approach of widget interaction and follow-up questions to express their intent more clearly. For example, P'03.Slack commented, ``I wanted to know if the elderly survived and I realized that the system took a guess. I then explicitly asked for greater than 65.'' 

We categorized $18.2\%$ of the utterances as failure cases because either the chatbot could simply not understand the utterance or it resulted in an incorrect system response that the participant could not correct. Some of these cases were due to insufficient information in the underlying data. For example, P'02.Slack asked, ``how long did it take to rescue the Titanic survivors ?'' The chatbot could not resolve `rescue' to any analytical concept or data and simply showed the total number of Titanic survivors as its response. Other cases failed because the chatbot could not recognize certain tokens and concepts in the utterances. For example, ``which wineries would you suggest for a good cab? [P'08.Slack]'' 
resulted in no meaningful system response as the chatbot failed to understand that `cab' was the short form for `cabernet' and was unable to interpret `suggest.' When the chatbot prompted the participant to rephrase their query, the participant was able to get a more satisfactory answer by typing, ``show me wineries with good cabernet.'' In general, the prompts and clarifications that the chatbot posed to the participants in the event of these unsuccessful interpretations, encouraged them to actively restate their original utterance or pivot to a new conversation topic.

\subsubsection{Echo Show and Echo Devices}
We combine the discussion for both of these prototypes as the main modality of interaction was voice and there were commonalities in participant feedback across the two platforms. In addition, we highlight the interaction differences that we found when participants used the touchscreen on the Echo Show device, compared to a headless Echo device.

All participants found the voice chatbots to be easy to interact with. P'14.EchoShow said, ``I can operate the bot without the use of the hands and without thinking about my grammar when I ask something.'' Participants also mentioned that they enjoyed the voice interaction and were often curious about the answers they were provided. P'27.Echo reacted, ``I sometimes like to ask my Alexa random questions and see what she does. Even though we are talking about data here, I was curious to see what she (the chatbot) would answer. I was pretty pleased when I asked where should I go to get the best rated wine and it responded with Lewis Winery in Napa Valley.'' 

Ten participants asked the Echo Show chatbot 121 utterances in total (Avg 11.2 utterances per session). Based on coding of the videos and the notes, utterances were manually classified into one of the following categories: 46.05\% Fact-finding, 25\% Comparisons across values for a given attribute, 18.42\% Grouping attributes, 6.58\% Filtering one or more values, and 3.95\% requesting information about the dataset. The rest of the utterances were either resetting the context of the conversation or asking for a deeper data insight.

Ten participants asked the Echo chatbot 116 utterances in total (Avg 10.4 utterances per session). Based on coding of the videos and the notes, a majority (94.8\%) of the utterances were manually classified as fact-finding, similar to Study 1 and in stark contrast with the Slack and Echo Show chatbots. Other types of utterances included Filter (3.45\%), and asking for more information about the dataset (1.72\%).

Other differences across the two platforms are documented below.

\paragraph{\textbf{Effect of device modality on conversational behavior}}
There were more occurrences of follow-up conversations with the Echo Show (38.16\% of the utterances; average number of conversation turns was $2.05$) when compared to the Echo (24.14\%; average number of conversation turns was $1.19$). The Echo Show touchscreen served as a scaffold for conversation turns, prompting the user with a list of follow up questions that they could select or verbalize. P'11.EchoShow explained, ``It's hard to keep all the options in my head if the chatbot just speaks to me. The screen gave me hints so I can figure out my next step.'' In contrast, participants found it more challenging to mentally maintain the context of a conversation thread in a voice-only interaction. P'26.Echo remarked, ``It's a lot easier to just ask single questions and get single responses back; kind of how I use my Alexa at home for the weather.'' Note that follow-up conversation was considerably lower with both voice chatbots as compared to Slack ($50.9\%$ follow-up utterances).

\paragraph{\textbf{Utility of follow-up prompts in the system responses}}
As expected, with the voice-only Echo chatbot, follow-up questions were asked verbally as that was the only mode of interaction. Surprisingly, most participants interacting with the Echo Show also chose to ask follow-up questions verbally ($89.7\%$ of the follow-up utterances) rather than interacting with the list of options provided on the touch screen. When participants were asked for their reason of choice, many of them simply found it more convenient to verbally ask the question, as the chatbot was in active listening mode (P'13.EchoShow, P'16.Echoshow, P'18.EchoShow). Other participants rationalized their behavior with the way they typically interact with voice-based chatbots at home as P'11.EchoShow described -- ``I use an Echo Show in my kitchen, but my hands are always messy. It's easier for me to just ask for a recipe or set a timer without having to walk over and touch the screen. I'm just used to that.'' 

\paragraph{\textbf{Types of utterances and failure cases}}
Similar to voice-only interaction in Study 1, most questions ($94\%$) that participants asked of the Echo chatbot were fact-finding or single-turn conversations.  With the Echo Show chatbot, having the visualizations available along with the verbal responses encouraged more variety in the types of intents they asked. $20.5\%$ and $16.6\%$ of the utterances were identified as being ambiguous in the Echo Show and Echo chatbots respectively. These numbers are lower than what we observed in the Slack chatbot ($39.5\%$), as participants tended to be more explicit and concise when verbally interacting with the chatbots. P'30.Echo commented, ``Voice is a bit dicey. I'm not going get complicated with it and keep my questions simple and to the point.'' 

We identified $24.6\%$ and $27.3\%$ of the utterances as failure cases in the Echo Show and Echo chatbot interaction respectively. Similar to the Slack chatbot, failure cases were due to insufficient information in the underlying data or an inability to recognize concepts in the utterances. Compared to Slack ($18.2\%$), the voice chatbots had more failure cases, due to difficulty recognizing proper names (e.g., names of wineries and passenger names) and certain accents.  In the Echo Show scenario, participants found it easier to select an alternative option on the screen and continue their interaction. In comparison, participants interacting with the Echo chatbot would either restate their question slowly or use alternative simpler language (e.g., asking for wineries in  ``France'' as opposed to in ``Bordeaux''), hoping for a more appropriate system response. 

\section{Discussion}
Our explorations of different analytical chatbot modalities revealed variations in people's behavior and expectations. Below, we first revisit our research questions and then discuss opportunities for future work.

\subsection{Revisiting the Research Questions}

\textbf{RQ1 - NL utterances: What are the characteristics of NL questions that users ask through text vs. voice? What types of ambiguous and underspecified questions do they ask with these modalities?}:
Observations from our studies found that while voice-only interaction placed a heavy emphasis on fact-finding, chatbots that could respond with both NL and charts, engaged users in richer analytical workflows involving multi-turn conversation and analytical operations such as grouping, filtering, and comparisons. Conversational affordances of threading and interactive widgets further prompted multi-turn conversational interaction. We observed ambiguity around fuzzy concepts such as ``How many of the survivors were \emph{young}'' and ``Did the \emph{richer} passengers have better chances of surviving?'' and intent such as ``Have there been people who paid too much in their class?''. 

\noindent\textbf{RQ2 - Response expectations: What would users expect as a reasonable response? When do users expect only a text or voice response? When do they want charts to be shown along with a text or voice response? What are users' expectations of the charts shown in response to NL questions?}:
Our studies identified many user expectations around chatbot interaction, as documented in Section~\ref{sec:expectations}. These ranged from simple operations like automatically filtering nulls (and making those system actions transparent) to more elaborate requirements such as providing context in the chatbot responses to confirm that the query was understood, remind the user what they asked, and facilitate next steps in the analytical workflow. The user expectations also point towards areas where future research is needed, such as automatically identifying breaks in conversational flow where the context should be reset. 

\noindent\textbf{RQ3 - Modalities for repair: When the result is unexpected, how do users expect to repair the system behavior?}:
Follow-up utterances for repair either reformulated a misunderstood query or revised the chart to continue the analysis. With the Slack chatbot in Study 2, participants also extensively used the widgets for repair. Widgets also offered a mechanism to rapidly explore variants of a chart to see different perspectives (i.e. by adjusting filters). While all participants appreciated having various ways to repair their input, feedback on the ``was this what you expected?'' buttons was mixed as it sometimes interrupted a user's natural workflow and forced an extra step in the repair process. In addition, UI widgets were seldom used in the Echo Show, despite supporting both visual and voice. This observation highlights the need to support repair and refinement in the modality that people are most familiar or comfortable with on a given platform, and which keep them in a natural workflow.

\subsection{Future Directions}




Analytical chatbots are a promising medium for human-data interaction. Findings from our preliminary studies open up interesting research directions to explore.

\paragraph{\textbf{Support for greater versatility in intent understanding}} Understanding user intent and providing relevant responses is important to any chatbot platform, but is particularly challenging for analytics. \change{Similar to general chatbot interfaces, analytical chatbots are expected to exhibit the Maxims of Quantity and Manner. However, the notion of relevance is more nuanced for analytical inquiry and there are are opportunities to develop techniques for deeply understanding analytical intent.} Participants especially wanted the system to better support intent around comparisons. For example, P05 stated in response to a bar chart shown for their question ``can you show the total \% of survivors age 20 and up?'' -- ``I'm more of a visual person and it makes it more challenging to see what I'm looking for. If there is already a dashboard that is interactive, I could ask a question and see how the dashboard would respond. For a discrete question, I would like to see the discrete response relative to the whole.'' Future studies should explore more deeply how analytical chatbots can adapt to a range of questions in the context of established visual analytics systems ~\cite{askdata,powerbi,ibmwatson}. Studies should also explore additional analytical capabilities and datasets. Fact-finding questions were prevalent across all the three study variants, especially with voice input; users appreciated additional context with the simple system responses. More work needs to be done to ascertain the kinds of context that are most appropriate and helpful, including external sources of information.

\paragraph{\textbf{Establish trust and transparency}} Utterances can be ambiguous and chatbots need to infer information and provide sensible responses. We found that establishing trust was critical to user engagement, \change{conforming with the Maxims of Manner and Quality}. It was helpful to describe the provenance of responses, with the underlying logic and any data transformations. P08 commented, ``she gave me the right number but then she qualified the response with first class.'' P09 said, ``Repeating the phrases from the question is useful to make sure that the system understood me. The value of repeating validates the quality and accuracy of the system response.'' Additionally, we need to design chatbots to gracefully handle requests that cannot be supported or are not understood. P07 commented, ``I was happy that it showed some representation of the data, even if not the requested viz type.'' The studies showed that follow-up questions and widgets were useful affordances to repair and refine system choices. We also found that predictability in the chatbot behavior for handling different types of analytical questions further enhanced people's trust in the systems. P27 said, ``At first it felt a bit intimidating to figure out what I can ask. I now know what to expect after asking a few questions and I feel comfortable poking into the data more.''  Along the lines of trust, the business logic and data for chatbot platforms commonly exist in the cloud. As we see the prevalence of these platforms for enterprise data exploration, privacy and security are important issues to address for supporting wider adoption. 

\paragraph{\textbf{Promote collaboration}} \change{Grice's Maxims collectively support cooperative conversation between humans and with a computer.} Chatbot and messaging platforms such as Slack and Teams provide support for collaborative participation in communities of practice or special interest groups. Our current set of studies focused on interaction behaviors between a human and the chatbot and we did not consider multi-person conversation. It would be useful to better understand collaboration patterns in the context of cooperative conversational behaviors around data and visual analysis.

\paragraph{\textbf{Understand social cues and expectations}} People often apply social heuristics to computer interactions, focusing on cues in language, intonation, and emotions expressed by the chatbot agent~\cite{nass:2001}. \change{These social behaviors help provide the necessary grounding for conversational understanding, supporting characteristics described by Maxims of Manner and Quantity.} Research supports the benefits of using anthropomorphic characteristics in human-robot interactions~\cite{don:1992} in encouraging more conversational interaction and enhancing a person’s ability to make precise assumptions on how that agent is likely to act based on its persona. While we did not explicitly present the chatbots with human-like attributes such as a name or an avatar, we found some evidence of participants anthropomorphizing the voice chatbots. For example, P'26.EchoShow described the chatbot's behavior as, ``That was a tricky question, but she did her best to find the answer for me'' while others expressed politeness during their interaction using words such as ``please'' and ``thanks, chatbot!'' Further exploration is needed to understand the effect of anthropomorphic agency on people's attitude, trust, satisfaction, and biases when conversing about data.

\paragraph{\textbf{Leverage context and situation}} Lastly, we did not consider situational context when designing these analytical chatbots. Contextual chatbots can ascertain a user's intent by location, time, or role \change{to stay both informative and relevant to the conversation. Situational context can further bolster an analytical chatbot's behavior based on the Maxims of Quantity and Relation.} For example, a smart assistant considers a user's location when asked whether it will rain today. Adding additional intelligence to provide data insights (e.g., sharing metrics on the latest sales data to a company executive) as well as learning user preferences over time for the types of data questions that are of interest and the types of responses that are preferred, can further improve the utility of analytical chatbots.

\change{To summarize, the user expectations that people had towards analytical chatbots generally conform to Grice's Maxims while conversing with data. However, the analytical task, platform, and mode of interaction provide additional challenges and opportunities for richer and nuanced ways of understanding and expressing intent. Future work would need to explore these research directions both across and within each of the four maxims. Further, the complexity and interplay between language and data could introduce new techniques and experiences for scaffolding analytical conversations.}

\section{Conclusion}
Participants' enthusiastic reactions to our analytical chatbot prototypes suggest that chatbots are a promising and approachable design approach for data analytics. 
Although existing interaction design guidelines for chatbots are generally applicable here, our studies identified additional principles inherent to data exploration. Our results suggested approaches to interpret intent and reveal variations in user behavior based on the modality and interface affordances. Users tended to ask fact-finding or simple analytic questions, often as single-turn conversations, when interacting via voice alone. Adding charts, together with voice or text interaction, encouraged multi-turn conversation and deeper analytical questions. Threading and widgets in our Slack prototype especially encouraged this sort of behavior. Preferred affordances for follow-up adjustments differed across the platforms, with voice prompts being the overall preferred approach for voice-based chatbots and widgets heavily used in the Slack chatbot.
Overall, these studies provide a better understanding of principles for designing analytical chatbots, highlighting the intricacies of language pragmatics and analytical complexities with the UI capabilities of the platform. We hope that others find value in our insights around the design of intelligent analytical chatbots and explore new research directions in conversational discourse behavior along with novel user experiences.

\bibliographystyle{ACM-Reference-Format}
\bibliography{references}

\end{document}